\documentstyle[12pt,epsf]{article}    % Simple LaTeX, use with DINA4 as follows
\newlength{\dinwidth}                       
\newlength{\dinmargin}                      
\def\setboxz@h{\setbox\z@\hbox}% forgotten from two following files
\def\wdz@{\wd\z@}% forgotten from two following files
\expandafter\chardef\csname pre amssym.def at\endcsname=\the\catcode`\@
\catcode`\@=11
\def\undefine#1{\let#1\undefined}
\def\newsymbol#1#2#3#4#5{\let\next@\relax
 \ifnum#2=\@ne\let\next@\msafam@\else
 \ifnum#2=\tw@\let\next@\msbfam@\fi\fi
 \mathchardef#1="#3\next@#4#5}
\def\mathhexbox@#1#2#3{\relax
 \ifmmode\mathpalette{}{\m@th\mathchar"#1#2#3}%
 \else\leavevmode\hbox{$\m@th\mathchar"#1#2#3$}\fi}
\def\hexnumber@#1{\ifcase#1 0\or 1\or 2\or 3\or 4\or 5\or 6\or 7\or 8\or
 9\or A\or B\or C\or D\or E\or F\fi}
\font\tenmsa=msam10
\font\sevenmsa=msam7
\font\fivemsa=msam5
\newfam\msafam
\textfont\msafam=\tenmsa
\scriptfont\msafam=\sevenmsa
\scriptscriptfont\msafam=\fivemsa
\edef\msafam@{\hexnumber@\msafam}
\mathchardef\dabar@"0\msafam@39
\def\dashrightarrow{\mathrel{\dabar@\dabar@\mathchar"0\msafam@4B}}
\def\dashleftarrow{\mathrel{\mathchar"0\msafam@4C\dabar@\dabar@}}

\def\ulcorner{\delimiter"4\msafam@70\msafam@70 }
\def\urcorner{\delimiter"5\msafam@71\msafam@71 }
\def\llcorner{\delimiter"4\msafam@78\msafam@78 }
\def\lrcorner{\delimiter"5\msafam@79\msafam@79 }
\def\yen{{\mathhexbox@\msafam@55 }}
\def\checkmark{{\mathhexbox@\msafam@58 }}
\def\circledR{{\mathhexbox@\msafam@72 }}
\def\maltese{{\mathhexbox@\msafam@7A }}
\font\tenmsb=msbm10
\font\sevenmsb=msbm7
\font\fivemsb=msbm5
\newfam\msbfam
\textfont\msbfam=\tenmsb
\scriptfont\msbfam=\sevenmsb
\scriptscriptfont\msbfam=\fivemsb
\edef\msbfam@{\hexnumber@\msbfam}

\def\widehat#1{\setbox\z@\hbox{$\m@th#1$}%
 \ifdim\wd\z@>\tw@ em\mathaccent"0\msbfam@5B{#1}%
 \else\mathaccent"0362{#1}\fi}
\def\widetilde#1{\setbox\z@\hbox{$\m@th#1$}%
 \ifdim\wd\z@>\tw@ em\mathaccent"0\msbfam@5D{#1}%
 \else\mathaccent"0365{#1}\fi}
\font\teneufm=eufm10
\font\seveneufm=eufm7
\font\fiveeufm=eufm5
\newfam\eufmfam
\textfont\eufmfam=\teneufm
\scriptfont\eufmfam=\seveneufm
\scriptscriptfont\eufmfam=\fiveeufm

\catcode`\@=\csname pre amssym.def at\endcsname
\expandafter\ifx\csname pre amssym.tex at\endcsname\relax
                                         \else \endinput\fi
\expandafter\chardef\csname pre amssym.tex at\endcsname=\the\catcode`\@
\catcode`\@=11
\newsymbol\gtrsim 1326
\newsymbol\lesssim 132E
\catcode`\@=\csname pre amssym.tex at\endcsname
\newcommand{\GeV}{\mbox{\rm ~GeV}}

\newcommand{\Rp}{\mbox{$\not \hspace{-0.15cm} R_p$}}
\newcommand{\PT}{\mbox{$\not \hspace{-0.15cm} P_{\perp}$}}
\newcommand{\pbi}{\mbox{\rm ~pb$^{-1}$}}
\begin{document}
\pagestyle{plain}  % No page numbers (avoids problems for proceedings volume)
\begin{flushleft}
{\tt Published in the
     Proceedings of the Workshop Future Physics at HERA } \\
{\tt Edited by G. Ingelman, A. De Roeck and R. Klanner} \hfill
{\tt September 1996} \\
\end{flushleft}

\vspace*{0.5cm}
\begin{center}  \begin{Large} \begin{bf}
                R-Parity Violating Supersymmetry at HERA \\
                \end{bf}  \end{Large}
                \vspace*{6mm}
\begin{large}
    E. Perez$^{a}$, Y. Sirois$^b$ \\
\vspace*{0.1cm}
\end{large}
    $^a$ CEN-Saclay, DSM/DAPNIA/SPP, Gif-sur-Yvette, France \\
    $^b$ LPNHE Ecole Polytechnique, IN2P3-CNRS, Palaiseau, France \\
\vspace*{0.4cm}
\begin{large}    
     H. Dreiner \\
\vspace*{0.1cm}
\end{large}
     Rutherford Appleton Laboratory, Chilton, Didcot, Oxon, United Kingdom \\
\end{center}
\begin{quotation}
\noindent
{\bf Abstract:}
The phenomenology and prospects for a discovery of R-parity violating
Supersymmetry at HERA is analysed. Emphasis is put on the direct resonant 
production of squarks by electron-quark fusion and {\it all possible} 
subsequent decay modes of the squarks are considered. In particular, the full 
consequences of the mixing in the supersymmetric gaugino-higgsino sector are 
taken into account. A rich phenomenology emerges for HERA which offers a unique
sensitivity to new R-parity violating couplings and good discriminating power
against free parameters of the theory.
\end{quotation}

%======================================================================
\section{Introduction}
%======================================================================

% Why SUSY

Supersymmetry (SUSY) which fundamentally links fermions and bosons is
likely to be chosen as an essential property of a true theory 
beyond the Standard Model (SM). Among the most compelling
arguments for a SUSY world are the fact that local supersymmetric
transformations are fundamentally related to generators of
space-time translation (hence necessarily incorporates 
gravity)~\cite{SUSYGRAV},
the possibility to ``explain'' the hierarchy between the electroweak
mass scale and the Grand Unification or Planck mass 
scale, and the stability of a softly broken SUSY which ``naturally''
avoids the arbitrary fine tuning of the parameters which is 
necessary in the SM~\cite{SUSYTUNE}.

% Why MSSM

A natural framework for SUSY searches is provided by the Minimal
Supersymmetric extension of the Standard Model (MSSM)~\cite{SUSYMSSM}
which has predictive power within a finite and well defined set of 
free parameters and has neither been proven nor falsified by experimental
observations.
The latter is a non-trivial status given the remarkable precision tests
of the SM at the LEP collider over the recent years. It might have
to do with the fact that quantum corrections due to the sparticles
which otherwise respect all gauge symmetries of the SM tend to be small 
rendering indirect observations difficult.
It is in addition possible that direct searches for particles of the 
minimal field representation offered by the MSSM have at least partly
failed because they were looking at the wrong phenomenology.

% Why Rp-violating SUSY

The most general Yukawa couplings in a supersymmetric theory which is 
gauge invariant and minimal in terms of field content can be 
written~\cite{WMSSMRP} in the compact formalism of the superpotential 
as $ W_{SUSY} = W_{MSSM} + W_{\Rp} $. 
The $W_{MSSM}$ contains terms which are responsible for the 
Yukawa couplings of the Higgs fields to ordinary fermions.
The $ W_{\Rp} $ is given by:
\begin{equation}
 \label{eq:Wpot}
  W_{\Rp} = \lambda_{ijk} L_i L_j \bar{E}_k
                  + \lambda'_{ijk} L_i Q_j \bar{D}_k
                  + \lambda''_{ijk} \bar{U}_i \bar{D}_j \bar{D}_k
\end{equation}
where $ijk$ are generation indices of the superfields $L, Q, E, D$ and $U$. 
The $L$ and $Q$ are left-handed doublets while ${\bar E}$, ${\bar D}$ 
and ${\bar U}$ are right-handed singlet superfields for charged leptons, 
down and up-type quarks, respectively. 
The $\lambda$ and $\lambda'$ terms induce lepton number violation while
the $\lambda''$ terms induce baryon number violation. 
% The first ($\lambda$) term allows to couple a slepton to a lepton and 
% a quark, the second ($\lambda'$) term couple a squark to a lepton and 
% a quark. Terms $\lambda''_{ijk} \bar{U}_i \bar{D}_j \bar{D}_k$ 
% violate baryonic number, and couple a squark to two quarks.
In the strict MSSM framework, one imposes that the SUSY theory be 
also minimal in terms of allowed couplings and all $W_{\Rp}$ terms are 
avoided by imposing a strict conservation of the R-parity defined as
$R_p = (-1)^{3B+L+2S} = 1$ (for particles) $= -1$ (for sparticles),
where $S$ denotes the spin, $B$ the baryon number and $L$ the 
lepton number. 
Imposing this discrete symmetry is a somewhat {\it ad hoc} prescription.
Another viable~\cite{ZNSYMM} 
% (e.g. for anomaly cancellation with the minimal field content of the MSSM) 
and less restrictive discrete symmetry is the
$B$-parity which imposes only baryon number conservation 
(i.e. $\lambda'' = 0$).
Vanishing $\lambda''$ couplings is sufficient to avoid unacceptable 
$n - \bar{n}$ oscillations and fast proton decay.
%YS Add references here ?
Moreover from the cosmological point of view, the observed 
matter/antimatter asymmetry imposes much more severe constraints 
on $\lambda''$ than on $\lambda$ or $\lambda'$~\cite{BARYON}.  
Finally, $B$-parity is favoured over $R$-parity conservation in a large
class of superstring inspired models~\cite{ZNSYMM}.
It is also interesting to note that the $\lambda$ and $\lambda'$ terms 
in~(\ref{eq:Wpot}) which have no equivalent in the SM arise in a fundamental 
way from the fact that $SU(2)$-doublet lepton superfields have the same gauge 
quantum numbers as the Higgs supermultiplets.

%
%   -> Why Rp-violating SUSY at HERA ?
%
The $ep$ collider HERA which provides both leptonic and baryonic 
quantum numbers in the initial state is ideally suited for \Rp\
searches. This was realized long ago and was first investigated  
theoretically in the context of the previous HERA 
Workshop~\cite{HERA91} which motivated early
experimental searches~\cite{H194}. 
The cases $\lambda' \neq 0$ which could lead to resonant production of 
squarks via $e$-$q$ fusion offers of course the
most exciting prospects.
Recent investigations~\cite{SUSY95,PEREZ96,H196}  have shown that a 
new and rich phenomenology (different for $e^-$ and $e^+$ beams) emerges 
when considering the full complexity of the mixing in the 
gaugino-higgsino sector of the theory.
This is studied in more details in this contribution in view of future
high luminosity runs at HERA.

The case of associated ${\tilde e}$-${\tilde q}$ production at HERA
followed by the \Rp-decay of the sfermions has already been
studied in detail and also in view of high luminosity runs at
HERA in~\cite{DREINERMORA}.
Via this process one can probe significantly smaller Yukawa couplings
than via the resonant production but only at smaller sfermion
masses.

%========================================================================
\section{Phenomenology of $\Rp$ SUSY}
%========================================================================

%-----------------------------------------
\subsection{Modelling and Free Parameters}
%-----------------------------------------
\label{sec:model}

The $\lambda'_{ijk} L_i Q_j \bar{D}_k$ terms in the \Rp\ extension of
the MSSM correspond in expanded field notation to the Lagrangian 
\begin{eqnarray}
 {\cal{L}}_{L_{i}Q_{j}\bar{D}_{k}} &=
 & \lambda^{\prime}_{ijk} \left[ -\tilde{e}_{L}^{i} u^j_L \bar{d}_R^k
   - e^i_L \tilde{u}^j_L \bar{d}^k_R - ({\bar{e}}_L^i)^c u^j_L
   \tilde{d}^{k*}_R \right.           \nonumber \\
\mbox{} &\mbox{}
 & \left. + \tilde{\nu}^i_L d^j_L \bar{d}^k_R + \nu_L \tilde{d}^j_L
          \bar{d}^k_R + ({\bar{\nu}}^i_L)^c d^j_L \tilde{d}^{k*}_R
\right]
  +\mbox{h.c.}
 \label{eq:LRP}
\end{eqnarray}
where the superscripts $^c$ denote the charge conjugate spinors and the
$^*$ the complex conjugate of scalar fields. 
Among the 27 possible $\lambda'_{ijk}$ couplings, the cases $i=1$ can
lead to direct squark resonant production and are thus of special
interest at HERA.
These cases are studied first in this paper assuming conservatively 
that one of the $\lambda'$ dominates. 

The masses of the scalar quarks and scalar leptons, bosonic sparticle 
partners of the SM fermions, are treated here as free parameters. 
In the gaugino-higgsino sector, there are four neutralinos $\chi^0_i$
($i=1 \dots 4$) which are mixed states of the photino $\tilde{\gamma}$, 
the zino $\tilde{Z}$ and the supersymmetric partners $\tilde{H^0_1}$ 
and $\tilde{H}^0_2$ of the two neutral Higgs fields.
Two charginos $\chi^{\pm}_j$ ($j=1,2$) are mixed states of the winos 
$\tilde{W}^{\pm}$ and of the SUSY partners of the charged Higgs fields.
The masses and couplings of the $\chi^0$ and $\chi^{\pm}$ are calculated
in terms of the MSSM basic parameters~:  
\begin{itemize}
 \item $M_1$ and $M_2$, the $U(1)$ and $SU(2)$ soft-breaking 
       gaugino mass terms; 
 \item $\mu$, the Higgs mixing parameter;
 \item $\tan \beta$, the ratio of the vacuum expectation values of the 
       two neutral Higgs fields.
\end{itemize}
The number of free parameters is reduced by assuming a relation
at the Grand Unification (GUT) scale between $M_1$ and $M_2$ (see 
Appendix for detail). No other GUT relations are used and in particular
the gluino ($\tilde{g}$) mass is left free.  

We moreover consider the following simplifying assumptions~:
\begin{itemize}
 \item all squarks (except the stop) are quasi-degenerate in mass; 
 \item the lightest supersymmetric particle (LSP) is the lightest
       neutralino $\chi_1^0$;
 \item gluinos are heavier than the squarks such that decays
       $\tilde{q} \rightarrow q + \tilde{g}$
       are kinematically forbidden.
\end{itemize}
It should be made clear that there are no compelling cosmological
constraints in \Rp\ models which impose that the LSP be neutral and
colourless. 
Other possible choices for the LSP (e.g. $\tilde{g}$ or $\chi^{\pm}$) would 
not significantly change the search and analysis strategy and will only be 
briefly discussed.
In \Rp\ models, in contrast to the strict MSSM, the
LSP is generally unstable. 
This leads to event topologies which differ strongly
% wildly 
from the characteristic ``missing energy'' signal due to LSP's escaping 
detection in the MSSM.
% Supersymmetric particles have to be produced by pair in the MSSM
Hence, except for exclusion limits derived from indirect searches 
(e.g. from the intrinsic width of the $Z^0$), the mass constraints
obtained in the MSSM framework do not apply directly in \Rp\ models.
The search for $\Rp$ squarks is ``complementary'' (hence mandatory) 
to that performed in the strict MSSM framework.

%---------------------------------
\subsection{Squark Production}
\begin{figure}[htb]
  \vspace{-0.5cm}
  \begin{center}
   \mbox{\epsfxsize=0.7\textwidth \epsffile{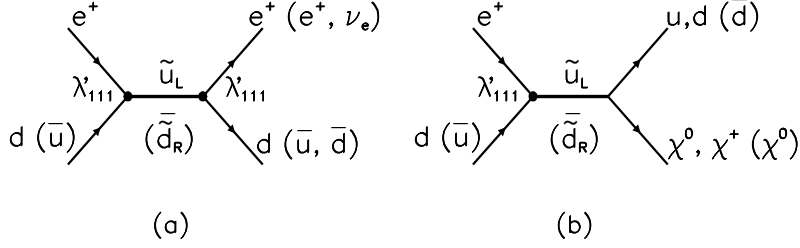}}
  \end{center}
  \vspace{-0.3cm}
  \caption[]{ \label{fig:sqdiag}
  {\it \Rp\ resonant production of $\tilde{u}_L$ or $\tilde{d}_R$ squarks
       in $e^+p$ collisions with subsequent (a) \Rp\ decay or (b) gauge
       decay involving a (generally) unstable gaugino-higgsino ($\chi^0$ 
       or $\chi^+$). }}
\end{figure}

%---------------------------------
%
The resonant squark production mode through direct $e$-$q$ fusion 
is illustrated in Fig.~\ref{fig:sqdiag} for $\lambda'_{111} \neq 0$.
%
%---------------FIGURE 1: Feynman diagrams -----------------------------
%-----------------------------------------------------------------------
%
By gauge symmetry, only $\tilde{u}_L$-like or $\tilde{d}_R$-like squarks
(or their charge conjugates) can be produced in $ep$ collisions.
The production of ``left'' squarks (i.e. supersymmetric partners of 
left-handed quarks) is the dominating process if HERA delivers 
positrons, since the fusion occurs via a $d$ valence quark. 
On the contrary, with electrons in the initial state, 
mainly ``right'' squarks are produced. 
This dichotomy has important consequences since ``left'' and ``right'' 
squarks have different allowed or dominant decay modes as will be
seen in the following sections.
In particular, new exotic final state topologies might have sizeable
contributions in $e^+p$ collisions.

HERA offers a high sensitivity to any of the nine $\lambda'_{1jk}$ 
couplings, in contrast to most indirect processes.
The production processes allowed for each $\lambda'_{1jk}$ 
are listed in Table~\ref{tab:sqprod} for an $e^+$ beam. 
%
% --- TABLE 1 : PRODUCTION PROCESSES  ----------------------
%           e+ beam
\begin{table*}[b]
  \renewcommand{\doublerulesep}{0.4pt}
  \renewcommand{\arraystretch}{1.2}
 \begin{center}
 \begin{tabular}{p{0.45\textwidth}p{0.55\textwidth}}
         \caption
%YS      {\small
         {\it \label{tab:sqprod}
         Squark production processes at HERA ($e^+$ beam)
         via a R-parity violating
         $\lambda^{'}_{1jk}$ coupling.} &

   \begin{tabular}{|c|c|c|}
   \hline 
   $\lambda^{'}_{1jk}$ & \multicolumn{2}{c|}{Production processes}
\\
   \hline
   \vspace{-3mm}
   \mbox{} & \mbox{} & \mbox{} \\
   111 & $e^+ +\bar{u} \rightarrow \bar{\tilde{d}}_R$
       &$e^+ +d \rightarrow \tilde{u}_L $\\
   112 & $e^+ +\bar{u} \rightarrow \bar{\tilde{s}}_R$
       &$e^+ +s \rightarrow \tilde{u}_L $\\
   113 & $e^+ +\bar{u} \rightarrow \bar{\tilde{b}}_R$
       &$e^+ +b \rightarrow \tilde{u}_L $\\
   121 & $e^+ +\bar{c} \rightarrow \bar{\tilde{d}}_R$
       &$e^+ +d \rightarrow \tilde{c}_L $\\
   122 & $e^+ +\bar{c} \rightarrow \bar{\tilde{s}}_R$
       &$e^+ +s \rightarrow \tilde{c}_L $\\
   123 & $e^+ +\bar{c} \rightarrow \bar{\tilde{b}}_R$
       &$e^+ +b \rightarrow \tilde{c}_L $\\
   131 & $e^+ +\bar{t} \rightarrow \bar{\tilde{d}}_R$
       &$e^+ +d \rightarrow \tilde{t}_L $\\
   132 & $e^+ +\bar{t} \rightarrow \bar{\tilde{s}}_R$
       &$e^+ +s \rightarrow \tilde{t}_L $\\
   133 & $e^+ +\bar{t} \rightarrow \bar{\tilde{b}}_R$
       &$e^+ +b \rightarrow \tilde{t}_L $\\
   \hline 
  \end{tabular}
  \end{tabular}
  \end{center}
\end{table*}
%           e- BEAM
%
% \begin{tabular}{p{0.45\textwidth}p{0.55\textwidth}}
%         \caption
%         {\small \label{tab:sqprod}
%         Squarks production processes at HERA ($e^+$ beam)
%         via a R-parity violating
%         $\lambda^{'}_{1jk}$ coupling.} &
%
%   \begin{tabular}{|c|c|c|}
%   \hline 
%   $\lambda^{'}_{1jk}$ & \multicolumn{2}{c|}{Production processes}
%\\
%   \hline
%   \vspace{-3mm}
%   \mbox{} & \mbox{} & \mbox{} \\
%   111 & $e^- + u \rightarrow {\tilde{d}}_R$
%       &$e^- + \bar{d} \rightarrow \bar{\tilde{u}}_L $\\
%   112 & $e^- + u \rightarrow {\tilde{s}}_R$
%       &$e^- + \bar{s} \rightarrow \bar{\tilde{u}}_L $\\
%   113 & $e^- +u \rightarrow {\tilde{b}}_R$
%       &$e^- +\bar{b} \rightarrow \bar{\tilde{u}}_L $\\
%   121 & $e^- +c \rightarrow {\tilde{d}}_R$
%       &$e^- +\bar{d} \rightarrow \bar{\tilde{c}}_L $\\
%   122 & $e^- +c \rightarrow {\tilde{s}}_R$
%       &$e^- +\bar{s} \rightarrow \bar{\tilde{c}}_L $\\
%   123 & $e^- +c \rightarrow {\tilde{b}}_R$
%       &$e^- +\bar{b} \rightarrow \bar{\tilde{c}}_L $\\
%   131 & $e^- +t \rightarrow {\tilde{d}}_R$
%       &$e^- +\bar{d} \rightarrow \bar{\tilde{t}}_L $\\
%   132 & $e^- +t \rightarrow {\tilde{s}}_R$
%       &$e^- +\bar{s} \rightarrow \bar{\tilde{t}}_L $\\
%   133 & $e^- +t \rightarrow {\tilde{b}}_R$
%       &$e^- +\bar{b} \rightarrow \bar{\tilde{t}}_L $\\
%   \hline 
%  \end{tabular}
%  \end{tabular}
%----------------------------------------------------------------------
Squark production via $\lambda'_{1j1}$ is especially interesting in 
$e^+p$ collisions as it involves a valence $d$ quark, 
whilst $\lambda'_{11k}$ are best probed with an $e^-$ beam since
squark production then involves a valence $u$ quark.

\begin{figure}[htb]
  \begin{center}
    \begin{tabular}{p{0.45\textwidth}p{0.55\textwidth}}
      \vspace{-4.0cm}
      \caption[]{ \label{fig:xsect}
      {\it Squark production cross-sections in $e^+p$ collisions for 
           a coupling $\lambda'_{1j1}=0.1$. }} &
      \mbox{\epsfxsize=0.5\textwidth \epsffile{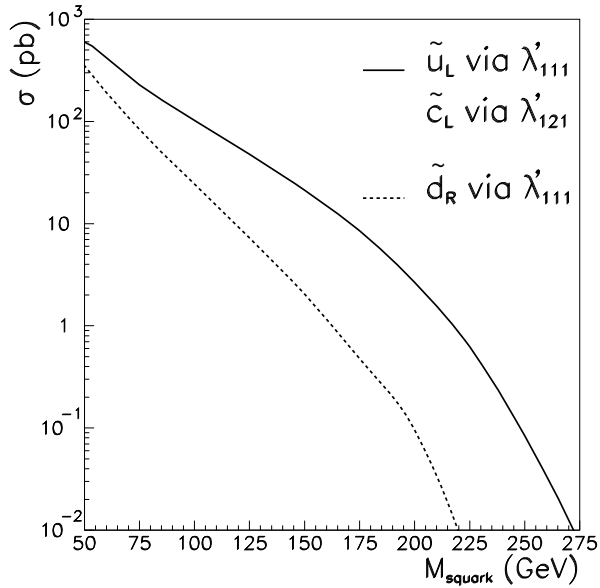}}
    \end{tabular}
  \end{center}
\vspace{-1.0cm}
\end{figure}

Figure~\ref{fig:xsect} shows the production cross-sections via 
$\lambda'_{111}$ for $\tilde{u}_L$ and $\bar{\tilde{d}}_R$,  and
for $\tilde{c}_L$ via $\lambda'_{121}$, each plotted for coupling
values of $\lambda'=0.1$. 
In the narrow width approximation, these cross-sections are simply
expressed as 
\begin{equation}
   \sigma_{\tilde{q}} = \frac{\pi}{4 s}
                        \lambda'^2 q'(\frac{M^2}{ s }) 
\end{equation}
where $\sqrt{s} = \sqrt{ 4 E^0_e E^0_p } \simeq 300 \GeV$ is the
energy available in the CM frame for incident beam energies of
$E^0_e = 27.5 \GeV$ and $E^0_p = 820 \GeV$, and $q'(x)$ is the
probability to find the relevant quark (e.g. the $d$ for $\tilde{u}_L$
or $\tilde{c}_L$ and the $\bar{u}$ for $\bar{\tilde{d}}_R$) with
momentum fraction $ x = M^2 / s \simeq M^2_{\tilde{q}} / s$ in the proton.
Hence the production cross-section approximately scales in $\lambda'^2$. 
The full kinematic domain can be probed at HERA for couplings weaker
than the electromagnetic coupling (i.e. $\lambda^2/4\pi < \alpha_{em}$)
given an integrated luminosity of $\simeq 500 {\mbox{pb}}^{-1}$.
%
% --- FIGURE 2: Production cross-section ------------------------------
%
%-----------------------------------------------------------------------

%YS  Introduce here a statement on relative X-sections for ~u_L and
%YS  ~d_R with e- and e+ beams ... justifies partly both beams

%--------------------------------
\subsection{Squark Decays}
%--------------------------------

The squarks decay either via their Yukawa coupling into ordinary
matter fermions, or in a first step via their gauge coupling into 
a quark and a neutralino $\chi_i^0$ ($i=1 \ldots 4$) or a chargino 
$\chi_j^{+}$ ($j=1,2$).
The former modes are henceforward called ``squark \Rp\ decays'' and
the latter ``squark gauge decays''.

\noindent
{\bf $\Rp$ decays of squarks:} \\
%    ------------------------
\noindent
In cases where both production and decay occurs through a
$\lambda'_{1jk}$ coupling (e.g. Fig.~\ref{fig:sqdiag}a for
$\lambda'_{111} \ne 0$), the squarks behave as scalar
leptoquarks~\cite{H1LQ95,BUCHMULL}.
For $\lambda'_{111} \ne 0$, the $\bar{\tilde{d}}_R$ resembles 
on event-by-event the
$\bar{S^0}$ leptoquark and decays in either $e^+ + \bar{u}$ or
$\nu_e + \bar{d}$  while the $\tilde{u}_L$ resembles the
$\bar{\tilde{S}}_{1/2}$ and only decays into $e^+ \bar{d}$.
The partial decay width reads~:
\begin{eqnarray}
  \Gamma_{\tilde{q} \rightarrow \Rp}
   = \Gamma_{{\tilde{u}_L} \rightarrow e^+ d}
   = \Gamma_{\bar{\tilde{d}_R} \rightarrow e^+ \bar{u} }
   = \Gamma_{\bar{\tilde{d}_R} \rightarrow \nu \bar{d} }
   = \frac{1}{16\pi} \lambda^{\prime 2}_{111} M_{\tilde{q}}
% \nonumber
\end{eqnarray}
so that squark $\Rp$ decays will mainly contribute at high mass
for large Yukawa coupling values $\lambda'$. 
Hence, the final state signatures consist of a lepton and a jet and
are, event-by-event, indistinguishable from the SM neutral
(NC) and charged current (CC) deep inelastic scattering (DIS).
% YS Make a quantitative statement on the narrowness of the
% YS resonance ..

\noindent
{\bf Gauge decays of squarks:} \\
%    ------------------------
\noindent
The MSSM Lagrangian contains terms coupling a sfermion to an ordinary
fermion and a gaugino-higgsino. 
The partial widths for squark gauge decays depend on MSSM parameters
via the composition of the neutralinos or charginos. 

Both $\tilde{q}_L$ and $\tilde{q}_R$ squarks can decay via 
$\tilde{q} \rightarrow q \chi_i^0$.
The partial width of the $\tilde{q} \rightarrow q \chi_i^0$ decay
is calculated to be
\begin{equation}
 \label{eq:f5}
    \Gamma_{\tilde{q} \rightarrow q + \chi^0_i} =
      \frac{1}{8\pi}(A^{2}+B^{2}) M_{\tilde{q}}
      \left(1-\frac{M^2_{\chi^0_i}}{M^2_{\tilde{q}}}\right)^2 
\end{equation}      
where~:
\begin{equation}
   A  = \frac{ g M_q N_{i4}} {2 M_W \sin \beta} \qquad , \qquad
   B  = e e_q N'_{i1} +
        g (0.5-e_q \sin^2 \theta_W) \frac{N'_{i2}}{\cos \theta_W} ,
\end{equation}        
and where $N_{ij}$ ($N'_{ij}$) is the transport matrix which
diagonalizes the neutralino mass matrix (see Appendix for detail) in the 
$\tilde{A}-\tilde{W}_3$ ($\tilde{\gamma}-\tilde{Z}$ basis). 
In practice, the $\chi_i^0$ masses and the exact values of the ``chiral'' 
couplings $A$ and $B$ depend on the relative $\tilde{\gamma}$, $\tilde{Z}$ 
and $\tilde{H}$ components of the $\chi_i^0$.

\begin{figure}[htb]
  \begin{center}
    \begin{tabular}{cc}
      \mbox{\epsfxsize=0.5\textwidth \epsffile{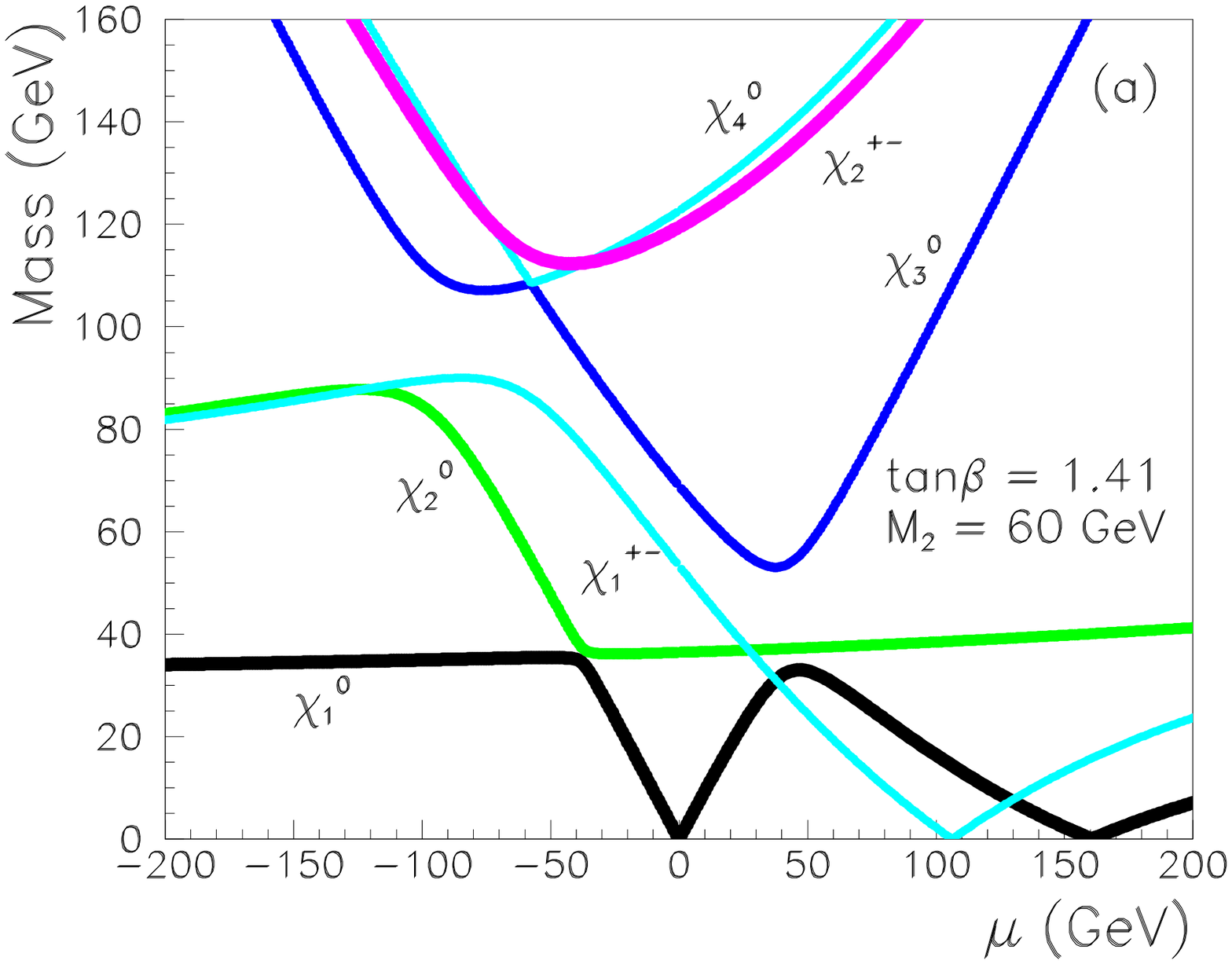}}
       &
      \mbox{\epsfxsize=0.5\textwidth \epsffile{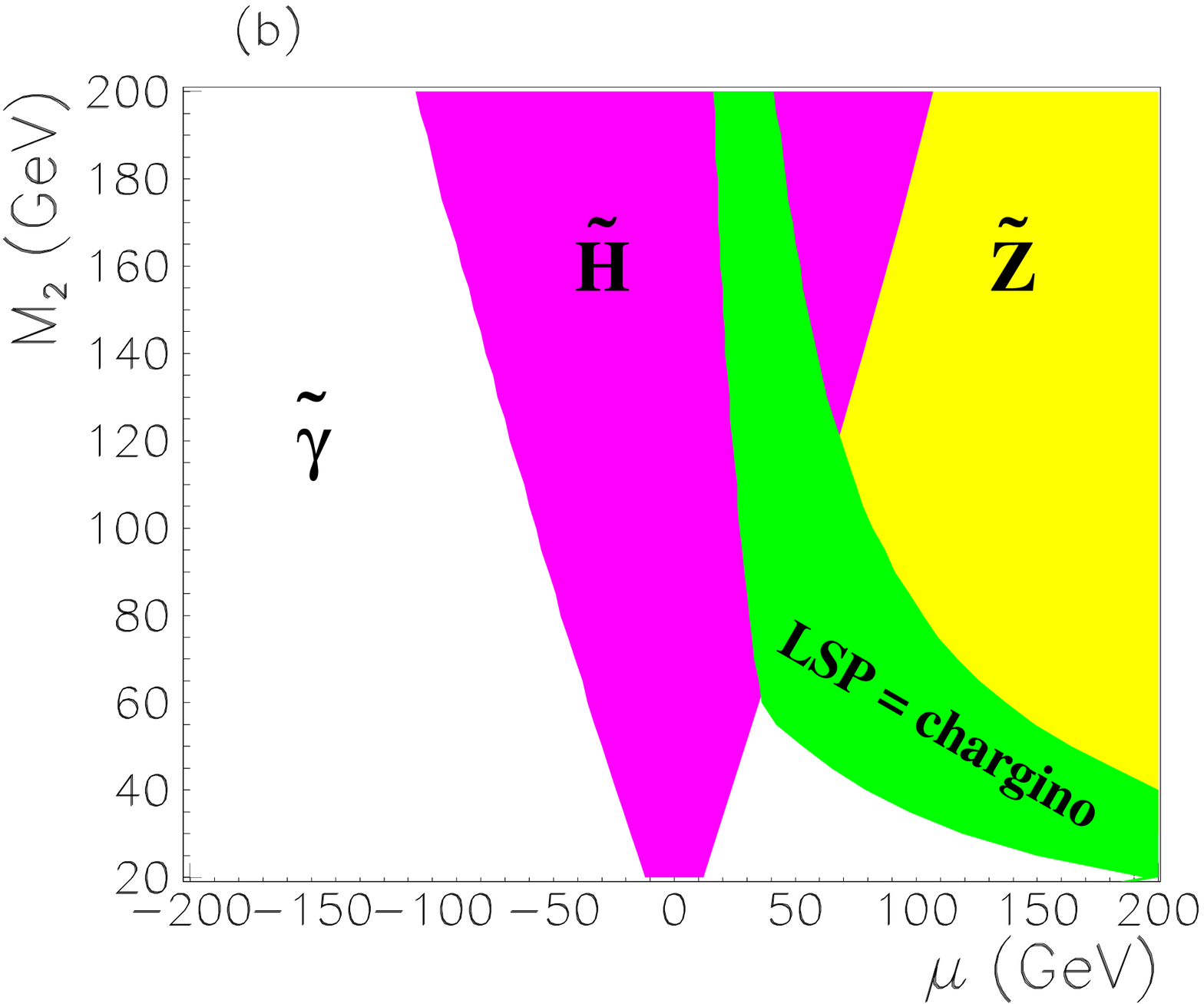}}
    \end{tabular}
    \caption[]{ \label{fig:massnlsp}
              {\it 
               (a) Physical masses of the $\chi_i^0$ and $\chi_i^{\pm}$
                   as a function of $\mu$ for $\tan \beta = 1$ and
                   $M_2 = 60 \GeV$; 
               (b) Main component of the LSP for $\tan \beta = 1$ }}
  \end{center}
\end{figure}

The dependence of the $\chi_i^0$ mass on the $\mu$ parameter is shown
in Fig.~\ref{fig:massnlsp}a for fixed $M_2$ and $\tan \beta$.
The dominant component ($\tilde{\gamma}, \tilde{Z}$ or $\tilde{H}$) of 
the lightest state $\chi_1^0$ is shown in Fig.~\ref{fig:massnlsp}b.
%
%---------------FIGURE 11: Gaugino-higgsino mass and LSP TOF ------------
%
%-----------------------------------------------------------------------
More details on the way the nature (and masses) of the various neutralino 
states depend on the basic MSSM parameters $M_2$, $\mu$ and
$\tan \beta$ can be found in the Appendix.

For a $\tilde{\gamma}$-like LSP, i.e. a $\chi_1^0$ dominated by its 
photino component, the $\tilde{q}$ to 
$q + \tilde{\gamma}$ coupling is proportional to the $q$ electric 
charge and the $\tilde{q}$ partial width reduces to 
\begin{equation}
  \Gamma_{\tilde{q} \rightarrow q + \tilde{\gamma}} =
     \frac{1}{8\pi}e^{2}e_q^2 M_{\tilde{q}}
     \left(1-\frac{M^2_{\tilde{\gamma}}}{M^2_{\tilde{q}}}\right)^2 \, .
\end{equation}     
In such a case, more than $90\%$ of the $\tilde{q} \rightarrow q \chi_i^0$ 
decays will involve the $\chi_1^0$.
A similar partial branching ratio holds for a $\tilde{H}$-like LSP with a
relatively large $\tilde{Z}$ component (e.g. in the $\tilde{H}$ region
close to the $\tilde{Z}$ region in Fig.~\ref{fig:massnlsp}b).
For a $\tilde{Z}$-like LSP, this branching ratio reduces to 
$20\% < {\cal{B}} < 80\%$.
Decays involving the LSP are negligible only in the $\tilde{H}$ domain
extending to negative $\mu$'s adjacent to the $\tilde{\gamma}$ domain
(Fig.~\ref{fig:massnlsp}b).
 
\begin{figure}[htb]
  \begin{center}
    \begin{tabular}{p{0.44\textwidth}p{0.56\textwidth}}
      \vspace{-4.0cm}
      \caption[]{ \label{fig:brsuldom}
      {\it Dominant gauge decay of a $150 \GeV$ $\tilde{u}_L$ squark. }}
      &
      \mbox{\epsfxsize=0.5\textwidth \epsffile{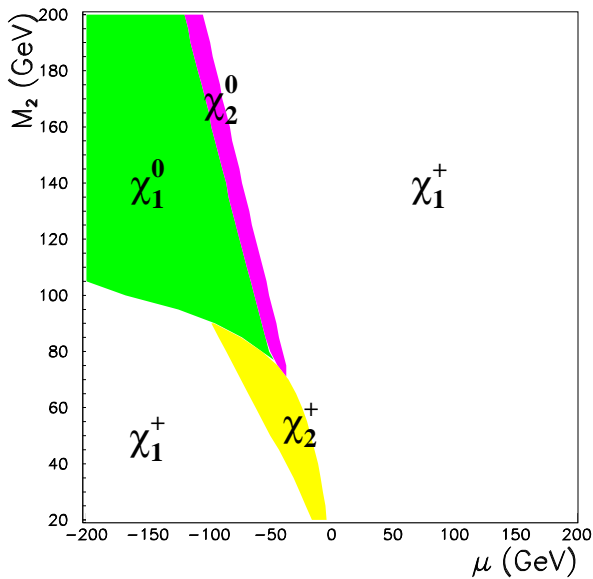}}
    \end{tabular}
  \end{center}
\end{figure}

(Almost) only the ${\tilde{q}}_L$ are allowed by gauge symmetry 
to decay into  $q' \chi_i^+$. 
This is because the $SU(2)_L$ symmetry which implies in the SM
that the right handed fermions do not couple to the $W$ boson
also forbids a coupling of $\bar{\tilde{q}}_R$ to the $\tilde{W}$.
The $\bar{\tilde{q}}_R$ decays involving the chargino is only possible 
through the $\tilde{H}^+$ component of the $\chi^+$ in which case the 
coupling is proportional to the $q'$ mass.
% Hence, the $\bar{\tilde{d}}_R$ can only weakly couple (in proportion
% to the $d$ quark mass) to the $\chi_j^+$ through its higgsino component.
%
Hence the decay $\bar{\tilde{q}}_R \rightarrow q' \chi_i^+$ is strongly 
suppressed for a $\bar{\tilde{q}}_R$ of the first or second generation.
The partial width of the $\tilde{q} \rightarrow q \chi_i^+$ decay
is obtained from~(\ref{eq:f5}) with the interchange 
$M_{\chi^0_i} \rightarrow M_{\chi^+_i}$ and with~:
% is calculated to be
% \begin{equation}
%   \Gamma( \tilde{q} \rightarrow \chi^{\pm}_i + q') =
%     \frac{1}{8\pi}(A^{2}+B^{2})g^{2}M_{\tilde{q}}
%     \left(1-\frac{M^2_{\chi^{\pm}_i}}{M^2_{\tilde{q}}}\right)^2 
% \end{equation}  
% where~:
\begin{equation}
   A = \frac{gV_{i1}} {\sqrt{2}} \qquad \qquad
   B = \frac{-gM_{q'} U_{i2} } {2 M_W \cos \beta} \,\, .
\end{equation}   
The regions of the $M_2$ vs $\mu$ plane where the $\tilde{u}$ decays
involving a chargino dominate are shown in Fig.~\ref{fig:brsuldom}.
%
%---------------FIGURE 3: Dominant gauge decays of squarks -------------
%-----------------------------------------------------------------------
In most of the parameter space, the $\tilde{u}_L$ squarks will mainly 
undergo a decay involving a chargino if kinematically allowed. \
The mass dependence of the $\chi_i^+$ states on the $\mu$ parameter is 
shown in Fig.~\ref{fig:massnlsp}a for fixed $M_2$ and $\tan \beta$.

%-----------------------------------------------
\subsection{Decays of the LSP}
\begin{figure}[htb]
\vspace{-0.6cm}
  \begin{center}
   \mbox{\epsfxsize=0.8\textwidth \epsffile{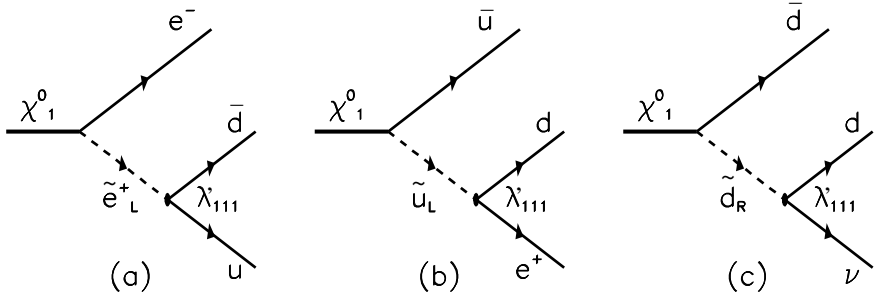}}
  \end{center}
 \vspace{-0.6cm}
 \caption[]{ \label{fig:lspdecay}
 {\it Example diagrams of the LSP decays $\chi_1^0 \rightarrow l q q'$ 
      involving a $\Rp$ Yukawa coupling. }}
\end{figure}

\begin{figure}[htb]
  \begin{center}    
    \begin{tabular}{cc}
       \mbox{\epsfxsize=0.5\textwidth \epsffile{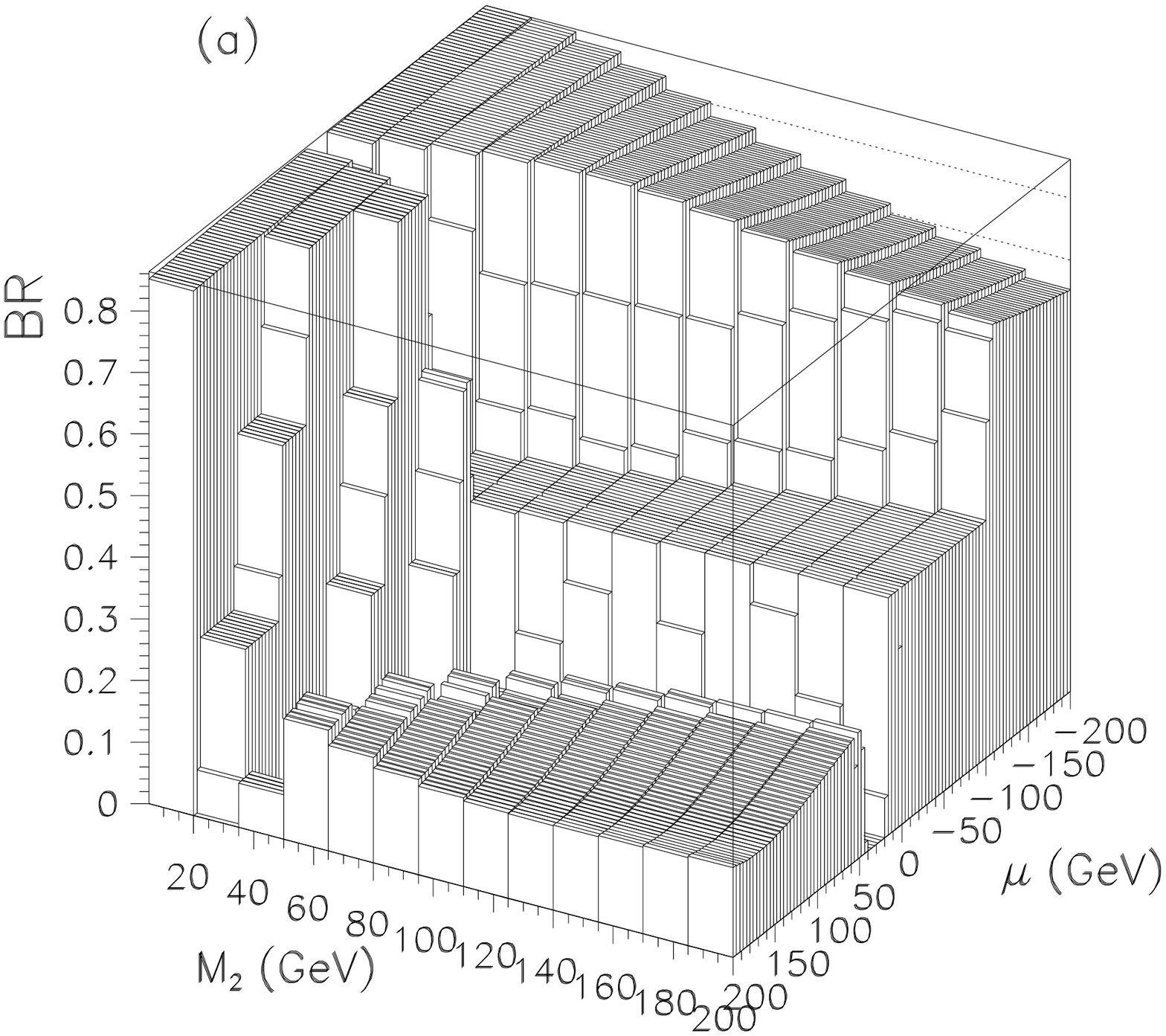}}
      & 
       \mbox{\epsfxsize=0.5\textwidth \epsffile{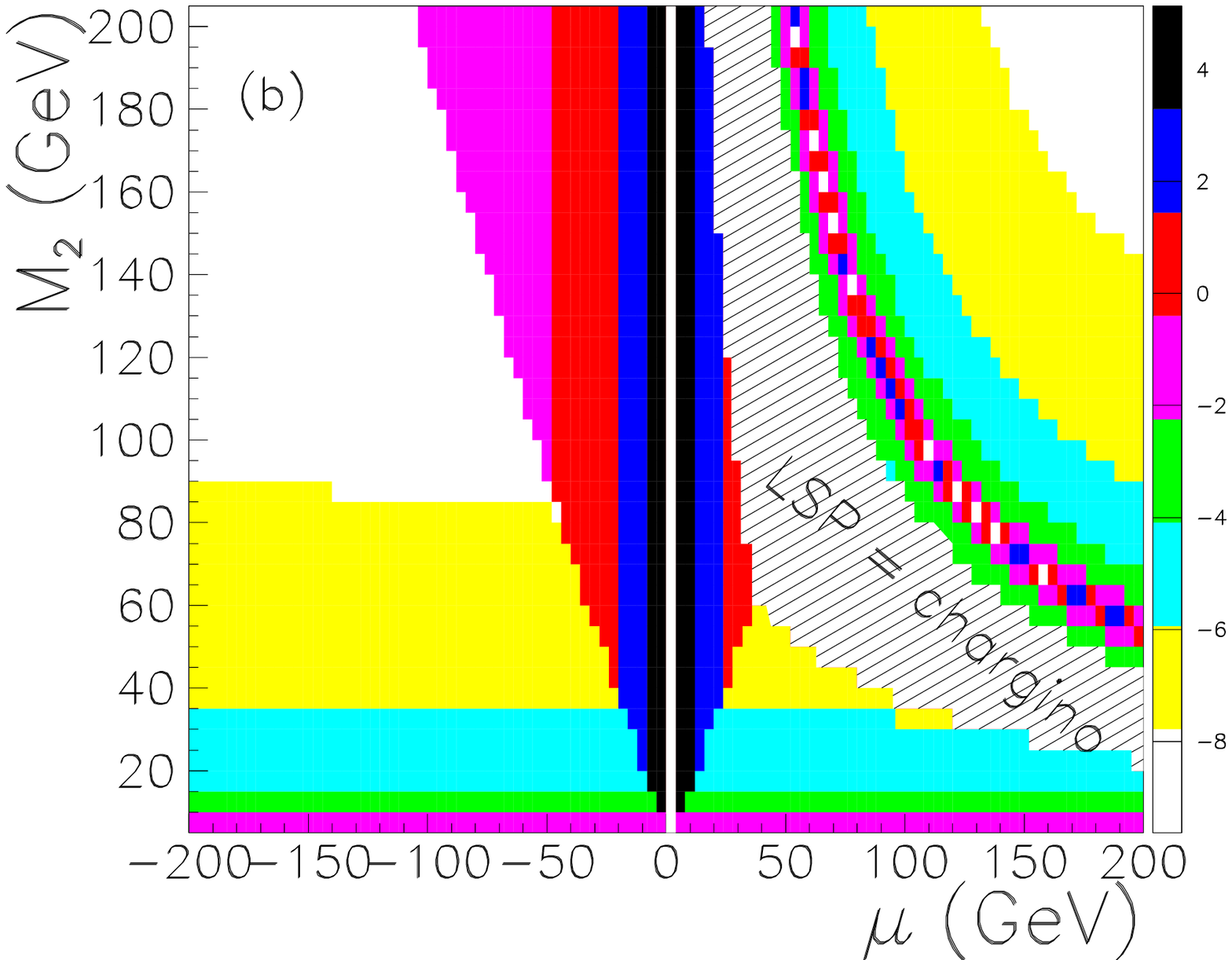}}
    \end{tabular}
    \caption[]{ \label{fig:lspbr}
    {\it (a) LSP ($\chi^0_1$) decay branching ratio into charged 
         leptons (i.e. $e^{\pm} + jets$), as a function of $\mu$ and 
         $M_2$ for sfermion masses $M_{\tilde{f}} = 150 \GeV$ and 
         $\tan \beta = 1$; 
         (b) $\log c \tau_0$ (m) of the LSP with $\lambda'=0.1$,
         the LSP mass is vanishingly small around $\mu =0$ and
         along the ridge at large $\mu + M_2$. }}
  \end{center}
\end{figure}

%-----------------------------------------------
%
In \Rp\ SUSY models with $\lambda'_{1jk} \neq 0$, the
LSP will undergo one of the following decays~:
$\chi^0_1 \rightarrow \nu \bar{d}_k d$,
$\chi^0_1 \rightarrow e^+ \bar{u}_j d_k$ or
$\chi^0_1 \rightarrow e^- u_j \bar{d}_k$.
%\begin{itemize}
% \item  $\chi^0_1 \rightarrow \nu \bar{d}_k d$ 
% \item  $\chi^0_1 \rightarrow e^+ \bar{u}_j d_k$ 
% \item  $\chi^0_1 \rightarrow e^- u_j \bar{d}_k$ 
%\end{itemize}
Representative diagrams of such decays are given in 
Fig.~\ref{fig:lspdecay}.
%
%---------------FIGURE 4: Examples decays of the LSP ----------
%
%-----------------------------------------------------------------------
%
The relevant matrix elements for these decays can be found 
in~\cite{DREINERMORA}.
They depend on the coupling $\lambda'$, but also on the parameters
$M_2$, $\mu$ and $\tan \beta$.
This dependence is illustrated in Fig.~\ref{fig:lspbr}a for the LSP decay 
$\chi^0_1 \rightarrow e^{\pm} q \bar{q}'$. 
Such decay modes are seen to be dominant ($63 \% < {\cal{B}}_R < 88 \%$) 
if the $\chi_1^0$ is $\tilde{\gamma}$-like in which case both the 
``right'' and the ``wrong'' sign lepton (compared to the incident
beam) are equally probable. 
This leads to largely background free striking signatures for lepton 
number violation.
The latter will dominate if the $\chi_1^0$ is $\tilde{Z}$-like.
A $\tilde{H}$-like $\chi_1^0$ will most probably
be long lived and escape detection since its coupling to
fermion-sfermion pairs is proportional
to the fermion mass~\cite{GUNION}.
This is illustrated in Fig.~\ref{fig:lspbr}b, which shows the flight
distance $c \tau_0$ of the $\chi^0_1$ in the plane $(M_2, \mu)$ for
$\lambda'=0.1$.  
The $c \tau_0$ exceeds 1 m in most of the $\tilde{H}$-like domain
surrounding the singularity at $\mu = 0$ where $M_{\chi_1^0} = 0$
at tree level.
Hence processes involving a higgsino-like $\chi_1^0$ will be
affected by an imbalance in transverse momenta.
%
%---------------FIGURE 5a: BR(LSP -> e+/- + jets)  -------------------
%---------------       5b: LSP flight distance and nature ------------
%-----------------------------------------------------------------------

%------------------------------------
\subsection{Decays of Charginos}
\begin{figure}[htb]
  \begin{center}
      \mbox{\epsfxsize=0.8\textwidth \epsffile{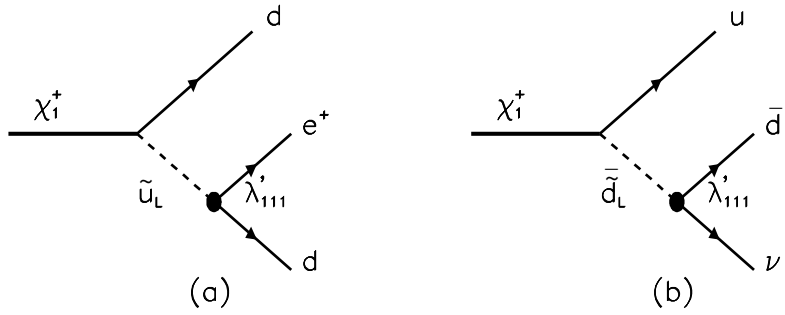}}
  \end{center}
 \vspace{-0.6cm}
 \caption[]{ \label{fig:charrp}
      {\it Representative diagrams for the $\chi^+_1$ decays 
      $\chi_1^+ \rightarrow l q q'$ involving a $\Rp$ Yukawa coupling. }}
\end{figure}

%------------------------------------
%
R-parity conserved $\chi^+$ decays into a $\chi^0$ and two matter 
fermions, have been investigated in detail in~\cite{BARTLLEP2}, where the
relevant matrix elements can be found. 
New decay modes of the $\chi^+$ into $e^+ + d_j + \bar{d}_k$ or 
$\nu_e + u_j +{\bar d}_k$ are allowed by the $\Rp$ couplings 
$\lambda'_{1jk}$ as illustrated in Fig.~\ref{fig:charrp}. 
%
%---------------FIGURE 6: Chargino decays -----------------------------
%
%-----------------------------------------------------------------------
%

The branching ratio of the $\chi_1^+$ into these $\Rp$ modes 
is obtained using the partial widths calculated from the relevant
matrix elements. 
Considering for instance the process 
$\chi^+ \rightarrow e^+ + d_j + \bar{d}_k$ (Fig.~\ref{fig:charrp}a) in 
the following notation
$\chi^+_1(k)\rightarrow e^+(l) + \bar{d}(q_1) + d(q_2)$,
and using Mandelstam variables $s=(q_1+q_2)^2=(k-l)^2$,      
$t=(k-q_1)^2=(l+q_2)^2$ and $u=(k-q_2)^2=(l+q_1)^2$,
the squared matrix element can be written as~:
\begin{eqnarray}
 |{\cal M}|^2_{\not
   R_p}=3g^2\lambda^{\prime2}_{111}|V_{11}|^2\left(\frac{s(M^2_{\chi}-s)}
   {|R(s)|^2}+\frac{t(M^2_{\chi}-t)}{|D(t)|^2}
   -{\cal R}e\frac{I(s,t,u)}{R(s)D(t)}\right)
\end{eqnarray}
where the propagators $R$ and $D$ and the interference term $I$ are~:
\begin{eqnarray}
R(s)&=&s-m^2_{{\tilde{\nu}}_L}, \\
D(t)&=&t-m^2_{{\tilde{u}}_L}, \\
I(s,t,u) &= &s(M^2_{\chi}-s) - u(M^2_{\chi}-u) + t(M^2_{\chi}-t).
\end{eqnarray}
The matrix element corresponding to the process 
$\chi^+_1\rightarrow \nu_e + u +\bar d$ 
(Fig.~\ref{fig:charrp}b) is deduced from the previous one with the 
following substitutions:
$ e\rightarrow \nu$, $\bar{d} \rightarrow u$, $d \rightarrow d$
and $|V_{11}|^2\rightarrow |U_{11}|^2$. 
%
%---------------FIGURE 7: Chargino decay branching ---------------------
%  
\begin{figure}[htb]
    \vspace{-1.5cm}
  \begin{center}
    \begin{tabular}{p{0.45\textwidth}p{0.55\textwidth}}
      \vspace{-4.5cm}
      \caption[]{ \label{fig:chardec}
      {\it $\Rp$ chargino decay branching ratio as a function of 
           $\lambda'_{111}$ and $\mu$, for $M_2 = 80 \GeV$,
           $\tan \beta = 1$ and sfermions masses $=150 \GeV$;
           the hatched domain corresponds to $\mu$ values for which
           $M(\chi^+_1) < M(\chi^0_1)$. }} &
      \mbox{\epsfxsize=0.5\textwidth \epsffile{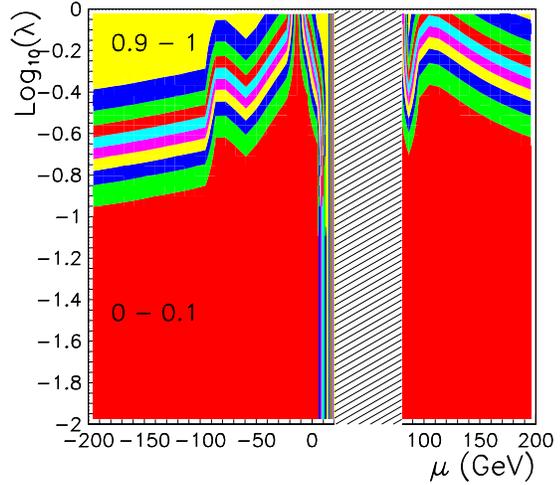}}
    \end{tabular}
  \end{center}
\end{figure}
%-----------------------------------------------------------------------
The corresponding partial width is obtained by integrating over phase
space as~:
\begin{equation}
  \label{eq:gamma}
   \Gamma = \int_{s=0}^{s=M^2_{\chi_1}}
            \int_{t=0}^{t=M^2_{\chi_1}-s}
             \frac{1}{M^3_{\chi^+_1}} \frac{1}{256 \pi^3}
                |{\cal{M}}|^2 ds dt  
\end{equation}                
The $\Rp$ decays of the $\chi^+_1$ will mainly dominate over MSSM decays
as soon as $\lambda'$ is not too small, as can be seen in
Fig.~\ref{fig:chardec}. For $\chi^0_1 \simeq \tilde{\gamma}$,
$\Rp$ decays of the chargino dominate over MSSM modes for coupling
values above $\simeq 0.25$, which is typically HERA's sensitivity
limit with current luminosity. 

\subsection{Classification of Final States}

Taking into account the dependence on the nature of the $\chi_1^0$,
the possible decay chains of the $\tilde{u}_L$ and $\tilde{d}_R$
squarks can be classified into eight distinguishable event
topologies listed in tables~\ref{tab:sqtopo1} and~\ref{tab:sqtopo2}
and labelled {\large S1} to {\large S8}. 
The {\large S1} and {\large S2} classes cover $\Rp$ squark decays.
The {\large S3} and {\large S4} classes are squark gauge decay 
topologies not accompanied by escaping transverse momenta $\PT$, 
while those with large $\PT$ are covered by classes 
{\large S5} to {\large S8}.
%
% --- TABLE 2: SUSY PROCESSES PART A  ----------------------------------
\begin{table}[htb]
 \renewcommand{\doublerulesep}{0.4pt}
 \renewcommand{\arraystretch}{1.0}
 \begin{center}
  \begin{tabular}{||c|c|l|l||}
  \hline \hline
  Channel  & $\chi_1^0$ & \multicolumn{1}{c|}{Decay processes}
                        & \multicolumn{1}{c||}{Signature} \\
           & nature & \multicolumn{1}{c|}{ }
                    & \multicolumn{1}{c||}{ }         \\ \hline
  S1 &  $\tilde{\gamma}$,$\tilde{Z}$,$\tilde{H}$
     &  \begin{tabular}{cccccc}
          $\tilde{q}$ & $\stackrel{\lambda'}{\longrightarrow}$
                      & $e^+$   & $q'$    &    &
        \end{tabular}
     &  \begin{tabular}{l}
        High $P_{\perp}$ $e^+$ + 1 jet
        \end{tabular} \\                                        \hline
%---------------------------------------------------------------------
  S2 &  \begin{tabular}{c}
          $\tilde{\gamma}$,$\tilde{Z}$,$\tilde{H}$ \\
          $\tilde{H}$
        \end{tabular}
     &  \begin{tabular}{cccccc}
          $\bar{\tilde{d}}_R$ & $\stackrel{\lambda'}{\longrightarrow}$
                      & $\nu_e$   & $\bar{d}$     &    & \\
          $\tilde{q}$         & $\longrightarrow$
                      & q         & $\chi_1^0$    &    &
        \end{tabular}
     &  \begin{tabular}{l}
         Missing $P_{\perp}$ + 1 jet
        \end{tabular} \\                                        \hline
%---------------------------------------------------------------------
  S3 & \begin{tabular}{c}
           $\tilde{\gamma}$,$\tilde{Z}$ \\ \\
           $\tilde{\gamma}$,$\tilde{Z}$,$\tilde{H}$ \\ \\
           $\tilde{\gamma}$,$\tilde{Z}$ \\ \\ \\ \\ \vspace{-0.5cm} \\
       \end{tabular}
     & \begin{tabular}{ccccll}
         $\tilde{q}$ & $\longrightarrow$
                     & $q$ & $\chi_1^0$  &  &  \\
         &  &        & $\stackrel{\lambda'}{\hookrightarrow}$
                     & $e^+ \bar{q}' q''$ & \\
         $\tilde{u}_L$ & $\longrightarrow$ & $d$ & $\chi_1^+$ &  & \\
         &  &        & $\stackrel{\lambda'}{\hookrightarrow}$
                     & $e^+ d \bar{d}$ &  \\
         $\tilde{u}_L$ & $\longrightarrow$
                     & $d$ & $\chi_1^+$ &  & \\
         &  &        & $\hookrightarrow$
                     & $W^+$ & $\hspace{-1.1cm}\chi_1^0$ \\
         &  &  &  & $\:|$
                  & $\hspace{-1.1cm}
                \stackrel{\lambda'}{\hookrightarrow}$ $e^+ \bar{q}' q''$
               \vspace{-0.2cm} \\
         &  &  &  & $\:\mid$ \vspace{-0.3cm} &  \\
         &  &  &  & $\:\rightarrow$  $q \:\: \bar{q}'$  &
       \end{tabular}
     & \begin{tabular}{l}
        High $P_{\perp}$ $e^+$ \\
        + multiple jets
       \end{tabular}\\                                          \hline
%---------------------------------------------------------------------
  S4 & \begin{tabular}{c}
          $\tilde{\gamma}$,$\tilde{Z}$ \\ \\
          $\tilde{\gamma}$,$\tilde{Z}$ \\ \vspace{-0.5cm} \\ \\ \\  \\
       \end{tabular}
     & \begin{tabular}{ccccll}
         $\tilde{q}$ & $\longrightarrow$
                     & $q$ & $\chi_1^0$ &  & \\
         &  &        & $\stackrel{\lambda'}{\hookrightarrow}$
         & $e^- \bar{q}' q''$ &  \\
         $\tilde{u}_L$ & $\longrightarrow$ & $d$ & $\chi_1^+$ &  & \\
         &  &        & $\hookrightarrow$
                     & $W^+$ & $\hspace{-1.1cm}\chi_1^0$ \\
         &  &  & & $\:|$
         & $\hspace{-1.1cm}
           \stackrel{\lambda'}{\hookrightarrow}$ $e^- \bar{q}' q''$
           \vspace{-0.2cm} \\
         &  &  &     & $\:\mid$ \vspace{-0.3cm}  &  \\
         &  &  &     & $\:\rightarrow$  $q \:\: \bar{q}'$  &
       \end{tabular}
     & \begin{tabular}{l}
        High $P_{\perp}$ $e^-$ \\ (i.e. wrong sign lepton) \\
        + multiple jets
       \end{tabular}\\
%---------------------------------------------------------------------
   \hline \hline
  \end{tabular}
  \caption[]
%         {\small 
          {\it \label{tab:sqtopo1}
               Squark decays in \Rp\ SUSY classified per
               distinguishable event topologies (PART I).
               The dominant component of the $\chi_1^0$ for which a
               given decay chain is relevant is given in the second
               column.
               The list of processes contributing to a given event
               topology is here representative but not exhaustive. }
%               e.g. the gauge decays of the $\chi_1^+$ involving a
%               virtual $W^+$ may also proceed via a virtual
%               sfermion as in Fig.~\ref{fig:sqdiag}b. }
 \end{center}
\end{table}
% --- TABLE 3: SUSY PROCESSES PART B  ----------------------------------
\begin{table}[htb]
 \renewcommand{\doublerulesep}{0.4pt}
 \renewcommand{\arraystretch}{1.0}
 \begin{center}
  \begin{tabular}{||c|c|l|l||}
  \hline \hline
  Channel  & $\chi_1^0$ & \multicolumn{1}{c|}{Decay processes}
                        & \multicolumn{1}{c||}{Signature} \\
           & nature & \multicolumn{1}{c|}{ }
                    & \multicolumn{1}{c||}{ }         \\ \hline
%
%---------------------------------------------------------------------
  S5 & \begin{tabular}{c}
          $\tilde{\gamma}$,$\tilde{Z}$ \\ \\
          $\tilde{\gamma}$,$\tilde{Z}$ \\ \vspace{-0.4cm} \\ \\ \\  \\
          $\tilde{\gamma}$,$\tilde{Z}$,$\tilde{H}$ \\ \\
          $\tilde{H}$ \\ \vspace{-0.6cm} \\ \\ \\
       \end{tabular}
     & \begin{tabular}{ccccll}
         $\tilde{q}$ & $\longrightarrow$
                     & $q$ & $\chi_1^0$ &  & \\
         &  &        & $\stackrel{\lambda'}{\hookrightarrow}$
                     & $\nu \bar{q}' q'$   &  \\
         $\tilde{u}_L$ & $\longrightarrow$
                     & $d$ & $\chi_1^+$   &  & \\
         &  &        & $\hookrightarrow$
                     & $W^+$ & $\hspace{-1.1cm}\chi_1^0$ \\
         &  &  & & $\:|$
         & $\hspace{-1.1cm}
           \stackrel{\lambda'}{\hookrightarrow}$ $\nu \bar{q}' q'$
           \vspace{-0.2cm} \\
         &  &  & & $\:\mid$ \vspace{-0.3cm} &  \\
         &  &  &     & $\:\rightarrow$  $q \:\: \bar{q}'$ & \\
         $\tilde{u}_L$ & $\longrightarrow$ & $d$ & $\chi_1^+$ &  & \\
         &  &        & $\stackrel{\lambda'}{\hookrightarrow}$
                     & $\nu u \bar{d}$ &  \\
         $\tilde{u}_L$ & $\longrightarrow$
                     & $d$ & $\chi_1^+$ &  & \\
         &  &        & $\hookrightarrow$
                     & $W^+$ & $\hspace{-1.1cm}\chi_1^0$ \\
         &  &  &     & $\hookrightarrow$ $q \:\: \bar{q}'$ &
       \end{tabular}
     & \begin{tabular}{l}
        Missing $P_{\perp}$ \\
        + multiple jets
       \end{tabular}\\                                          \hline
%---------------------------------------------------------------------
  S6 & \begin{tabular}{c}
          $\tilde{H}$ \\ \vspace{-0.1cm} \\ \\
       \end{tabular}
     & \begin{tabular}{ccccll}
         $\tilde{u}_L$ & $\longrightarrow$
                     & $d$ & $\chi_1^+$ &  & \\
         &  &        & $\hookrightarrow$
                     & $W^+$ & $\hspace{-1.1cm}\chi_1^0$ \\
         &  &  &     & $\hookrightarrow$  $l^+ \:\: \nu$  &
       \end{tabular}
     &
      \begin{tabular}{l}
         High $P_{\perp}$ $e^+$ or $\mu^+$ \\
         + missing $P_{\perp}$ + 1 jet
       \end{tabular}\\                                          \hline
%---------------------------------------------------------------------
  S7 & \begin{tabular}{c}
          $\tilde{\gamma}$,$\tilde{Z}$\\ \vspace{-0.5cm} \\ \\ \\ \\
       \end{tabular}
     & \begin{tabular}{ccccll}
         $\tilde{u}_L$ & $\longrightarrow$
                     & $d$ & $\chi_1^+$  &  & \\
         &  &        & $\hookrightarrow$
                     & $W^+$ & $\hspace{-1.1cm}\chi_1^0$ \\
         &  &  &  & $\:|$
         & $\hspace{-1.1cm}
           \stackrel{\lambda'}{\hookrightarrow}$ $e^{\pm} \bar{q}' q''$
          \vspace{-0.2cm} \\
         &  &  &     & $\:\mid$ \vspace{-0.3cm} &  \\
         &  &  &     & $\:\rightarrow$  $l^+ \:\: \nu$  &
       \end{tabular}
     & \begin{tabular}{l}
        High $P_{\perp}$ $e^{\pm}$ \\
        + high $P_{\perp}$ $e^+$ or $\mu^+$ \\
        + missing $P_{\perp}$ \\
        + multiple jets
       \end{tabular}\\                                          \hline
%---------------------------------------------------------------------
  S8 & \begin{tabular}{c}
          $\tilde{\gamma}$,$\tilde{Z}$\\ \vspace{-0.5cm} \\ \\ \\ \\
       \end{tabular}
     & \begin{tabular}{ccccll}
         $\tilde{u}_L$ & $\longrightarrow$
                     & $d$  & $\chi_1^+$  &  & \\
         &  &        & $\hookrightarrow$
                     & $W^+$ & $\hspace{-1.1cm}\chi_1^0$ \\
         &  &  &     & $\:|$
         & $\hspace{-1.1cm}
           \stackrel{\lambda'}{\hookrightarrow}$ $\nu \bar{q}' q'$
           \vspace{-0.2cm} \\
         &  &  &     & $\:\mid$ \vspace{-0.3cm} &  \\
         &  &  &     & $\:\rightarrow$  $l^+ \:\: \nu$   &
       \end{tabular}
     & \begin{tabular}{l}
        High $P_{\perp}$ $e^+$ or $\mu^+$ \\
        + missing $P_{\perp}$ \\
        + multiple jets
       \end{tabular}\\
%---------------------------------------------------------------------
   \hline \hline
  \end{tabular}
  \caption[]
%         {\small 
          {\it  \label{tab:sqtopo2}
               Squark decays in \Rp\ SUSY classified per
               distinguishable event topologies (Part II).
               As in table~\ref{tab:sqtopo1}, the list of processes
               given here is not exhaustive, e.g. the gauge decays
               $\chi_1^+ \rightarrow \chi_1^0 l^+ \nu$ and
               $\chi_1^+ \rightarrow \chi_1^0 q \bar{q}'$
               may also proceed via a virtual sfermion. }
 \end{center}

\end{table}
%
%-------------------------------------------------------------------------
A set of event selection cuts has been developed and discussed in
detail in~\cite{H194,H196}.

\begin{figure}[htb]
  \vspace{-1.0cm}
  \begin{center}
    \begin{tabular}{p{0.45\textwidth}p{0.55\textwidth}}
    \vspace{-4.0cm}
    \caption[]{ \label{fig:ysum}
     {\it Distribution of the variable $\Sigma y$ for neutral current DIS
         processes, and for a simulation of $75 \GeV$ squarks undergoing 
         gauge decays involving $20 \GeV$ neutralinos;
         the vertical line is the cut used in the H1 
         analysis~\cite{WARSAW96}. }} &
      \mbox{\epsfxsize=0.5\textwidth \epsffile{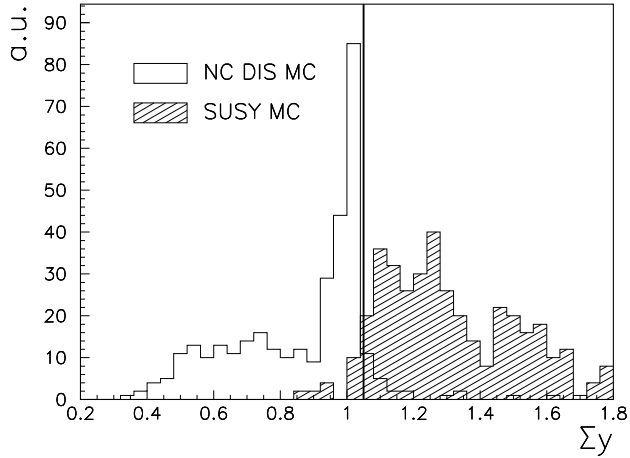}}
    \end{tabular}
  \end{center}
\end{figure}

For {\large S1} and {\large S3} (or {\large S4}), the DIS NC background is 
strongly suppressed by requiring a high $P_{\perp}$ $e^{\pm}$ found at 
high $y_e$, where $y_e = 1/2 (1 + \cos \theta_e^*)$ and $\theta_e^*$ is 
the electron angle in the $e-q$ CM frame.
The uniform decay of the scalar particle in the CM frame leads to a
flat $y_e$ spectrum for {\large S1} and one shifted towards largest
$y_e$ for {\large S3}. 
This is in contrast to the $1/y_e^2$ spectrum expected for the DIS NC 
background at fixed quark momentum fraction $x$.
For {\large S3} the H1 analysis~\cite{H196} has been 
improved~\cite{WARSAW96}, using $\theta^*$'s computed for the scattered 
electron and for the highest $P_{\perp}$ jet found in the azimuthal 
hemisphere opposite to the electron, and cutting on 
$\sum y = y_e + y_{jet}$ as shown in Fig.~\ref{fig:ysum}.
% This allows to reduce the background in this channel
% by a factor $\simeq 5$ in the low mass region ($M_{\tilde{q}}
% \leq 80 \GeV$). 
%
%---------------FIGURE 8: Ysum cut -------------------------------------
%
%-----------------------------------------------------------------------
Good signal detection efficiencies are obtained in each of these classes, 
reaching $\sim 70\%$ for {\large S1} and up to $\sim 60\%$ depending on 
$M_{\chi_1^0}$ for {\large S3}.

The {\large S4} topology with a wrong sign lepton in the final state is
quasi-background free. 
Event candidates in classes {\large S2} and {\large S5} to {\large S8} have
a large $\PT$. Classes {\large S2} and {\large S5} suffer from DIS CC
background and from tails of photoproduction background. 
The {\large S6} to {\large S8} topologies have one or many leptons in
the final states and are thus quasi-background free.
Typical signal detection efficiencies~\cite{H196} reach 
$\sim 30\% \rightarrow 80\%$ in these channels.

%---------------FIGURE 9: Branching ratios -------------------------------
%
\begin{figure}[b]
  \vspace{-2.0cm}
  \begin{center}
    \begin{tabular}{cc}
      \mbox{\epsfxsize=0.5\textwidth \epsffile{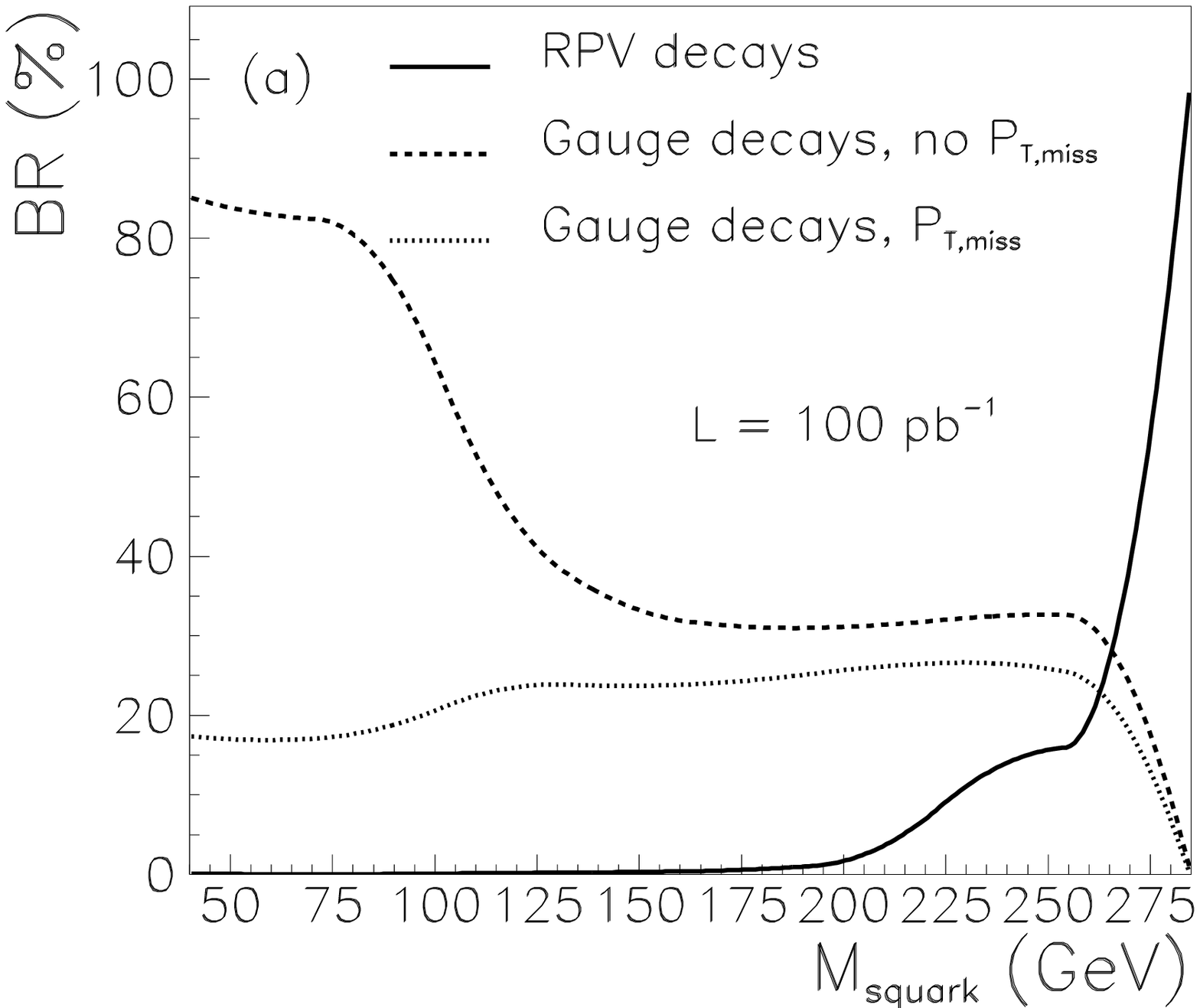}}
           &
%     \mbox{\epsfxsize=0.5\textwidth \epsffile{brcheckepsb.eps}}
      \mbox{\epsfxsize=0.5\textwidth \epsffile{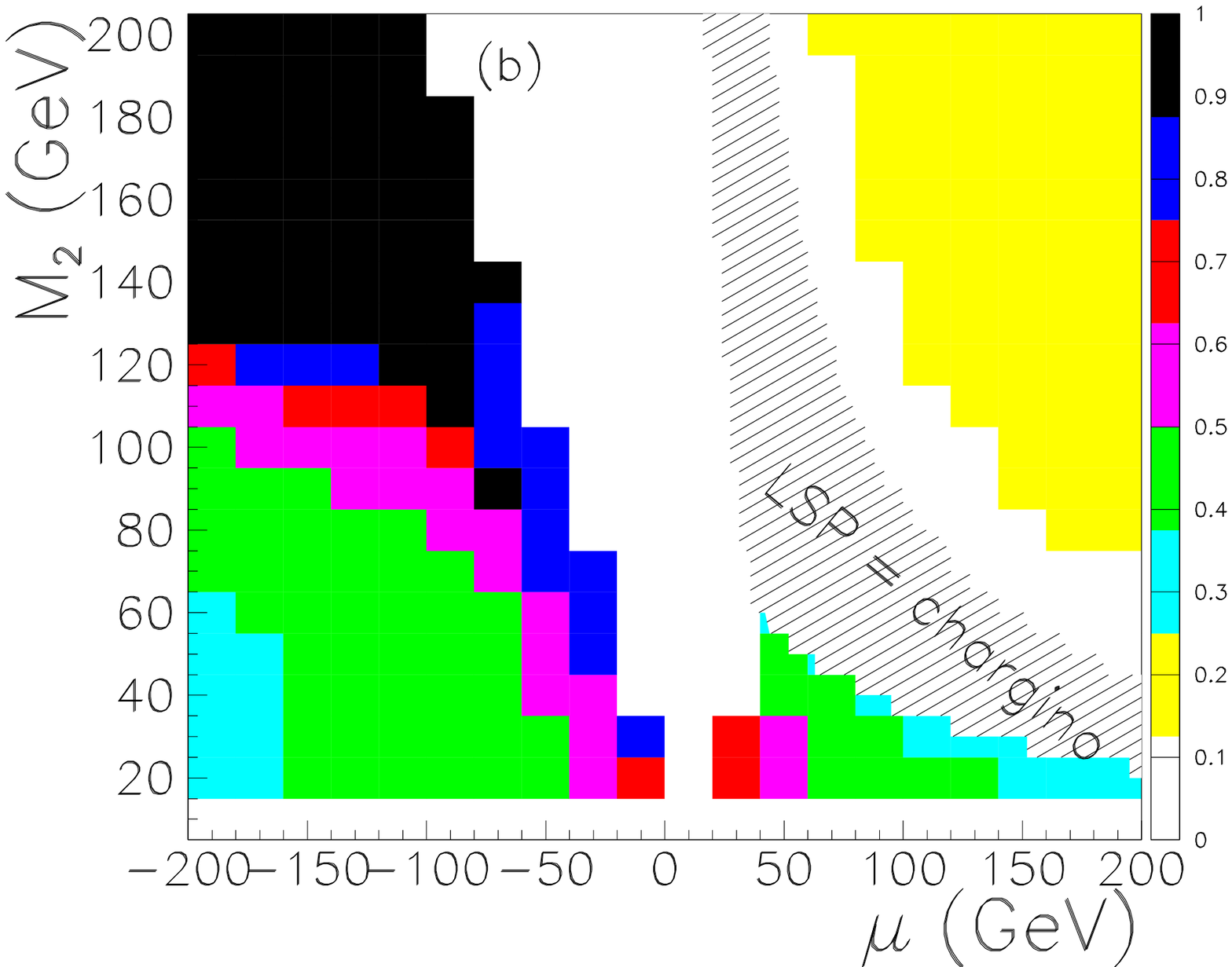}}
    \end{tabular}
  \end{center}
 \caption[]{ \label{fig:branching}
 {\it (a) Squark decay branching ratio as a function of squark mass integrated 
      over three distinct set of event topologies for 
      $\tan \beta = 1$ and $M_{\tilde{\gamma}} = 40 \GeV$.
%     (b) Convolution of the expected signal detection efficiency in H1
%     with the above branching ratio as a function of squark mass. }}
      (b) Ratio ${\cal{B}}_{S4}/{\cal{B}}_{S3}$ of the squark ``gauge'' decay 
      branching ratios without $\PT$ involving the like ($S3$) and 
      unlike ($S4$) sign lepton viewed in the $M_2$ versus $\mu$ plane;
      the plot is obtained for $M_{\tilde{q}} = 150 \GeV$ at the expected
      limit of $\lambda'$ coupling sensitivity for an integrated HERA
      luminosity of $100 \pbi$.}}
\end{figure}
%---------------------------------------------------------------------------
The relative contributions of the squark $\Rp$ and gauge decays are shown
in Fig.~\ref{fig:branching}a. Gauge decays are seen to dominate through most
of the accessible mass range. Only large Yukawa couplings can be probed at 
largest masses and thence $\Rp$ decays dominate.  
The shape of the curves in Fig.~\ref{fig:branching}a is only distorted
at lowish mass (e.g. $M_{\tilde{q}} \lesssim 75 \GeV$) 
when convoluting with signal detection efficiencies.  
The measurement of the relative branching ratio in $\large{S3}$ and 
$\large{S4}$ in case of a discovery, could be used to constrain the 
$\chi_1^0$ LSP nature in the MSSM parameter space as seen in 
Fig.~\ref{fig:branching}b.
%
%EP Higgsino-like => decay involving $\chi_1^0$ negligible .. 
%

% Relax hypothesis on gluino

It is interesting to note that the final state classification discussed
here should not be dramatically affected when relaxing the hypothesis
of section~\ref{sec:model}, e.g. in models where the $\tilde{g}$ are lighter 
than the $\tilde{q}$, or where the LSP is the $\chi_1^+$.
 
Assuming $M_{\tilde{g}} < M_{\tilde{q}}$, the decay
$\tilde{q} \rightarrow q + \tilde{g}$ will generally dominate.
If the $\tilde{g}$ is the LSP, the $\tilde{q}$ decay will be followed by 
the $\Rp$ decay $\tilde{g} \rightarrow q+q'+e^{\pm}$ or 
$\tilde{g} \rightarrow q+\bar{q} + \nu$.
In such a case, possible final states contain several jets and either
one electron or $\PT$.
These topologies correspond to channels {\large S3} and {\large S5}, 
previously considered. 
If $M_{\tilde{q}} > M_{\tilde{g}}$, with the LSP being the lightest
neutralino,
the $\tilde{g}$ arising from squark decay will undergo
$\tilde{g} \rightarrow q + \tilde{q}$, the latter squark being off-shell.
Possible final states are similar to those listed above, but more jets
would be expected. 
Assuming now that the LSP is the $\chi_1^+$ (see the relevant MSSM parameters
in Fig.~\ref{fig:massnlsp}b), a new event topology would only emerge for 
a relatively stable $\chi_1^+$ which could behave as a ``heavy muon''.
However, the time of flight of the $\chi^+_1$, obtained from
the integration~(\ref{eq:gamma}) over phase space, reads as~:
\begin{equation}
    \tau = \frac{4\pi}{g^2}  \frac{1}{|V_{11}|^2}
           \frac{1}{\lambda'^2} (8 \times 64 \pi^2)
           \left(  \frac{M_{\tilde{q}}}{M_{\chi^+_1}}  \right) ^4
           \frac{1}{M_{\chi_1^+}} 
\end{equation}            
which numerically leads to~:
\begin{equation}
   \tau = (2.5 \cdot 10^{-15} s)
         \left(  \frac{5. 10^{-3}}{\lambda'} \right)^2
         \frac{1}{|V_{11}|^2}
         \left( \frac{100 \GeV}{M_{\chi_1^+}}  \right)^5
         \left( \frac{M_{\tilde{f}}}{150 \GeV}  \right)^4   . 
\end{equation}
From this formula one obtains that the relevant parameter space
for the $\chi^+_1$ to decay outside the detector ($\gtrsim 1$m), 
is already excluded from the intrinsic
$Z^0$ width measurement at CERN~\cite{ALEPH}.

\section{Results for the Mass-Coupling Reach of HERA}
%        ===========================================

\begin{figure}[htb]
  \vspace{-1.0cm}
  \begin{center}
%    \begin{tabular}{p{0.4\textwidth}p{0.6\textwidth}}
    \begin{tabular}{cc}
%      \vspace{-5.0cm}
%      \caption[]{ \label{fig:l1j1lim}
%       {\it Exclusion upper limits at 95\% CL on the $\lambda'_{1j1}$
%            coupling as a function of squark mass which could be
%            reached with $e^+ p$ collisions at HERA 
%            ($\sqrt{s} \sim 300 \GeV$) for integrated luminosities of
%            $100 \pbi$ (dark shaded area) and $500 \pbi$ (shaded). }} 
%
      \mbox{\epsfxsize=0.45\textwidth \epsffile{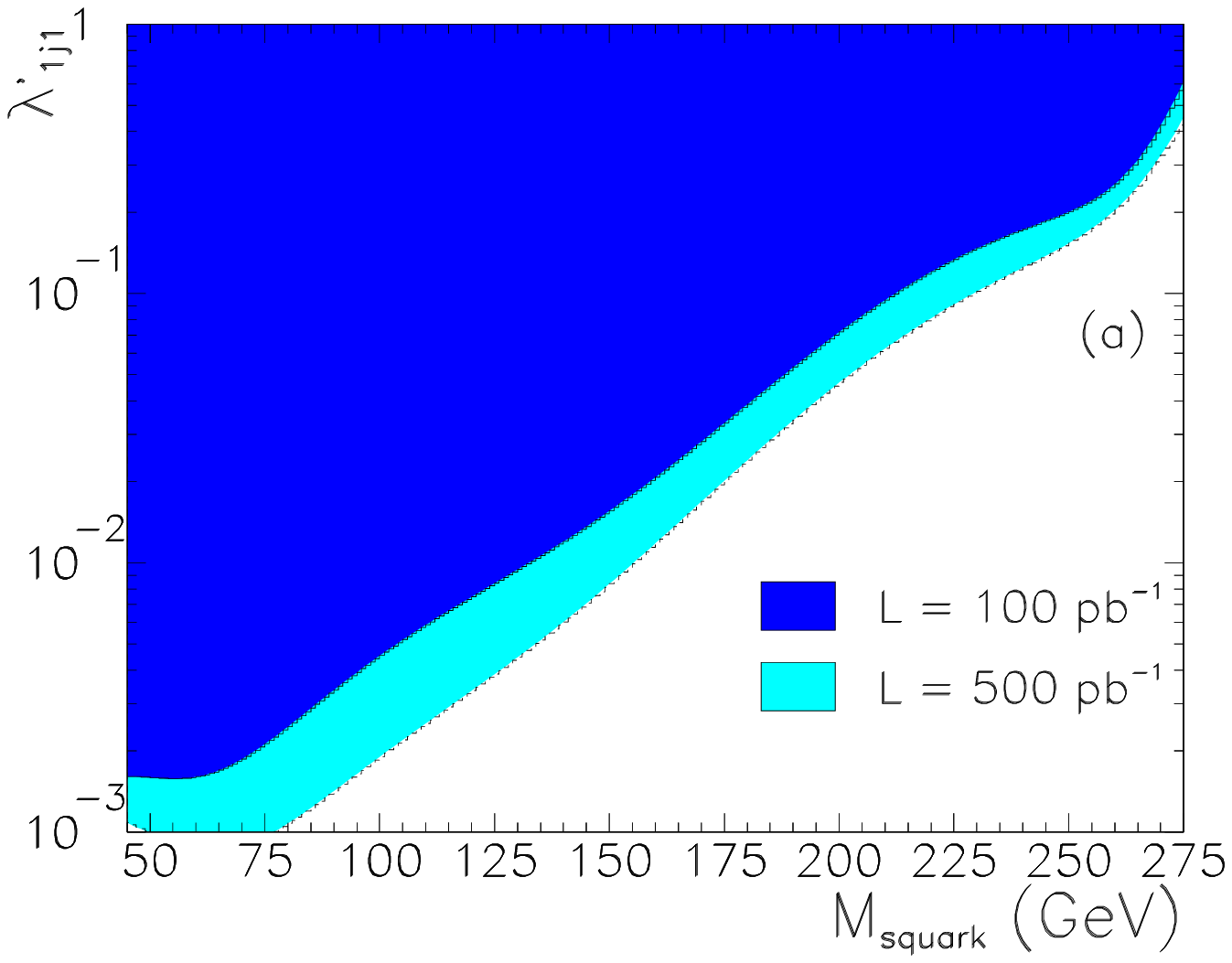}}
             &
      \mbox{\epsfxsize=0.65\textwidth \epsffile{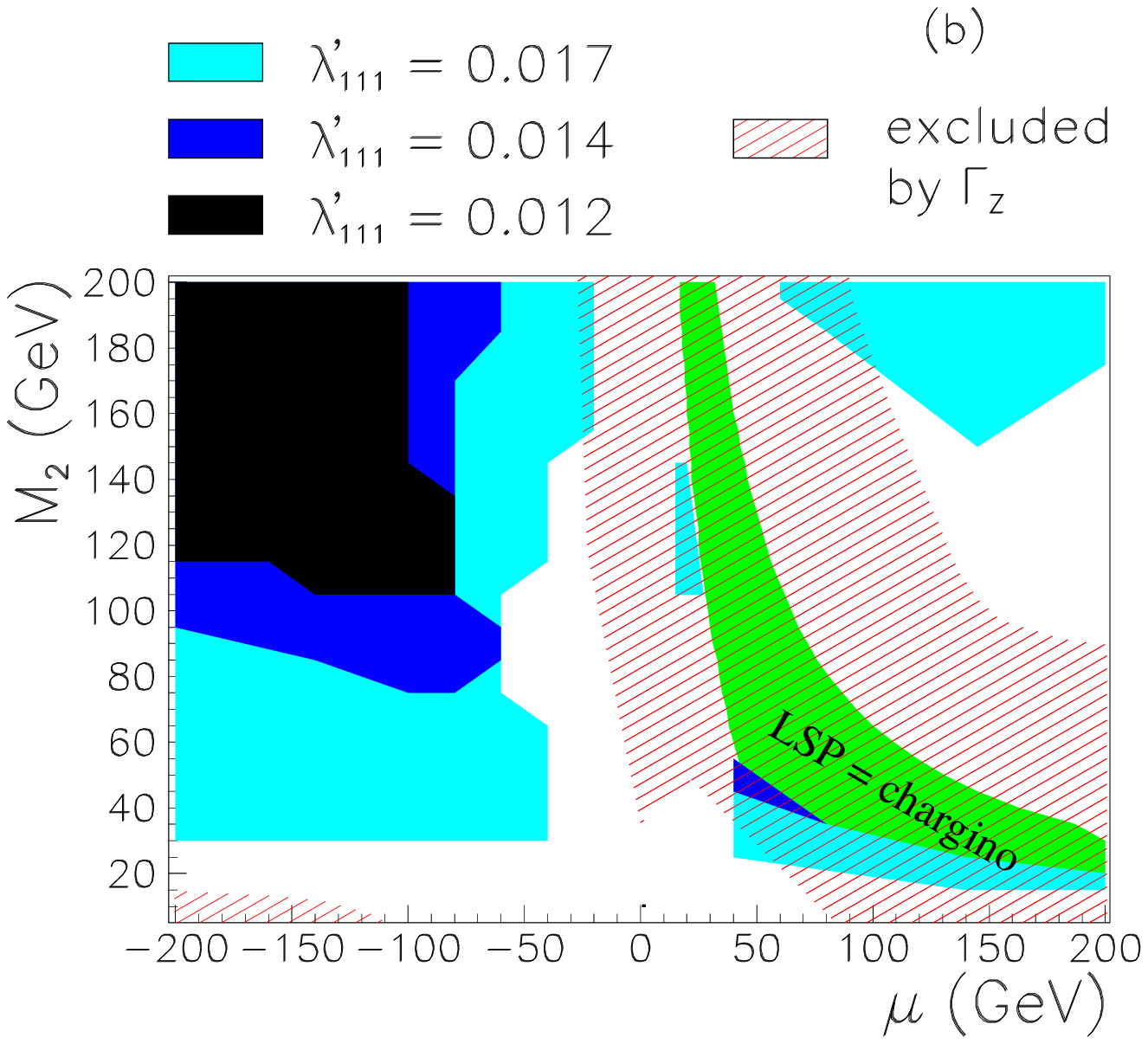}}
    \end{tabular}
  \end{center}
       \caption[xx]{ \label{fig:l1j1lim}
       {\it (a) Exclusion upper limits at 95\% CL on the $\lambda'_{1j1}$
            coupling as a function of squark mass which could be
            reached with $e^+ p$ collisions at HERA
            ($\sqrt{s} \sim 300 \GeV$) for integrated luminosities of
            ${\large L} = 100 \pbi$ (dark shaded area) and $500 \pbi$ (shaded);
            (b) Regions of the $M_2$ versus $\mu$ plane excluded for
            $L = 100 \pbi$ and for couplings $\lambda'_{1j1}$ equal or
            smaller to the exclusion upper limit at 
            $M_{\tilde{q}}=150 \GeV$. }}
\end{figure}

% 
In the absence of a significant deviation from the SM expectations, 
exclusion limits for the Yukawa couplings $\lambda'_{1jk}$
as a function of mass can be derived, showing the domain HERA could probe 
in the near future.
Results are shown for $\lambda'_{1j1}$ in Fig.~\ref{fig:l1j1lim} 
at $95 \%$ confidence level (CL), for integrated luminosities
$ {\large L} = 100 {\mbox{pb}}^{-1}$ and 
$ {\large L} = 500 {\mbox{pb}}^{-1}$.
These have been obtained assuming a $40 \GeV$ $\tilde{\gamma}$-like 
$\chi^0_1$, and combining all contributing channels.  
%
%---------------FIGURE 10: Exclusion limits -----------------------------
%
%----------------------------------------------------------------------
%
For ${\cal{L}} = 500 {\mbox{pb}}^{-1}$,
the existence of first generation squarks with \Rp\ Yukawa
coupling $\lambda'_{1j1}$ could be excluded for masses up to 
$\sim 270 \GeV $ for coupling strengths 
$\lambda'^2_{111} / 4\pi \, \gtrsim \alpha_{em} $. 

From the analysis of the $\lambda'_{1j1}$ case involving the
$\bar{\tilde{d}}_R$ and  $\tilde{u}_L$ squarks, limits can be
deduced on the $\lambda'_{1jk}$ by folding in the proper parton
densities. Such limits are given in Table~\ref{tab:limtab} at
$M_{\tilde{q}}=150 \GeV$ and for an integrated luminosity of
$ 500 {\mbox{pb}}^{-1}$. 
Also quoted in this table are the most severe existing 
indirect limits for each of these couplings. 
The most stringent concern couplings $\lambda'_{1jk}$ with
$j=k$ and come either from the non-observation of neutrinoless 
double-beta decay ($j=k=1$) or from constaints on the $\nu_e$ mass
($j=k=2,3$).
The limit from $ \beta \beta 0\nu$ decay depends on the gluino
mass and is given here for $M_{\tilde{g}}= 500 \GeV$.	
%
%====== TABLE: COMPARISONS WITH EXISTING INDIRECT LIMITS ==============
%  
\begin{table}[bt]                                                             
\setlength{\tabcolsep}{1.5pc}                                                   
\newlength{\digitwidth} \settowidth{\digitwidth}{\rm 0}                         
\catcode`?=\active \def?{\kern\digitwidth}                                      
% -----------------------------------------------------                         
\caption{{\it Exclusion upper limits at 95\% CL on the coupling 
         $\lambda'_{1jk}$ for $M_{\tilde{q}} = 150 \GeV$ and
         $M_{{\chi_1^0}} = 40 \GeV$ together with best existing
         indirect limits. The indirect limits have been scaled 
         from those found in the cited references to 
         $M_{\tilde{q}} = 150 \GeV$ and 95\% CL. }}                                 
\label{tab:limtab}                                                           
\begin{tabular*}{\textwidth}{@{}l@{\extracolsep{\fill}}rrrr}                    
\hline                                                                          
                 & \multicolumn{2}{l}{ {\bf HERA} sensitivity }                              
                 & \multicolumn{2}{l}{ Indirect limits} \\                      
\cline{2-3} \cline{4-5}                                                         
                 & \multicolumn{1}{r}{$\tilde{\gamma}$-like $\chi_1^0$}                                 
                 & \multicolumn{1}{r}{$\tilde{Z}$-like $\chi_1^0$}                                 
                 & \multicolumn{1}{r}{Value [Ref.]}                                 
                 & \multicolumn{1}{r}{Nature of the process}         \\                      
\hline                                                                          
   $\lambda'_{111}$  & 0.008  & 0.023  & 0.003~\cite{BETADECAY} & 
                                        $ \beta \beta 0\nu$ decay \\ 
   $\lambda'_{112}$  & 0.020   & 0.057  & 0.05~\cite{GIUDICE}  & 
                                          CC-universality      \\
   $\lambda'_{113}$  & 0.026   & 0.072  & 0.05~\cite{GIUDICE}  & 
                                          CC-universality      \\ 
   $\lambda'_{121}$  & 0.008   & 0.023 & 0.09~\cite{DAVIDSON}   & 
                                        Atomic Parity Viol. \\ 
   $\lambda'_{122}$  & 0.027   & 0.077  & 0.04~\cite{NUMASS}   & 
                                        $\nu_e$-mass \\ 
   $\lambda'_{123}$  & 0.043   & 0.012  & 0.5 ~\cite{BHATTA}  & 
                                        $ D^+ \rightarrow K $ decays \\ 
   $\lambda'_{131}$  & 0.007   & 0.024  & 0.09~\cite{DAVIDSON}   & 
                                        Atomic Parity Viol.        \\
   $\lambda'_{132}$  & 0.027   & 0.091  & 0.77~\cite{REEXP}   & 
                                        $R_e^{exp}$          \\ 
   $\lambda'_{133}$  & 0.068   & 0.230  & 0.0015~\cite{NUMASS} & 
                                        $\nu_e$-mass         \\
\hline                                                                          
\end{tabular*}                                                                  
\end{table}                                        
%

% COMPARISON WITH DIRECT LIMITS :

By the time HERA reaches high luminosity running conditions, 
new direct limits (or a discovery !) from other colliders will have 
further constrained the possible squark masses and SUSY parameters.
In $e^+e^-$ collisions, the direct squark pair production process does
not violate R-parity and LEP2 should directly probe squark 
masses up to $\sqrt{s}/2$, i.e. $\simeq 90 \GeV$.
In $p\bar{p}$ collisions, squarks can be produced in pair or in 
association with gluinos. 
No complete analysis in the $\Rp$-SUSY framework
has been performed yet with existing TEVATRON data.
Nevertheless, $\tilde{q}$ decay topologies similar to those described
here have been explored by D0~\cite{D0} and CDF~\cite{CDF} in 
scalar leptoquark or MSSM searches.
From these and from di-lepton data~\cite{ROY96}, one can infer that 
the range $200 \rightarrow 300 \GeV$ of $\Rp$-SUSY squark masses 
will most probably be not fully excluded by TEVATRON data for an
integrated luminosity of $\sim 100 \pbi$, thus leaving open a 
discovery window at HERA in the hypothesis 
$M_{\tilde{g}} \gg M_{\tilde{q}}$. 

%
% Lepton Flavor Violation:
%
If the presence of two simultaneously non-vanishing Yukawa couplings
(e.g. $\lambda'_{1jk}$ and $\lambda'_{ijk}$ with $i \neq 1$), resonant
$\tilde{q}$ production at HERA can be directly followed by a lepton flavor 
violation (LFV) decay leading to $\mu + {\mbox{jet}}$ or $\tau + {\mbox{jet}}$
signatures. 
Relevant analysis with existing data have been performed by the 
H1~\cite{H1LQ95} and ZEUS~\cite{ZEUSLFV} collaborations and limits comparable 
to the best existing indirect LFV limits have been derived in the context of 
$\Rp$-SUSY for a pure $\tilde{\gamma}$ like LSP. A new range of possible
coupling products could be probed with increasing luminosity~\cite{SCIULLI}.
 
%===========================================================================
\section{Summary and Conclusions}
%===========================================================================

The HERA potential for $R$-parity violating supersymmetry searches was
studied.
Direct resonant production of squarks through nine new Yukawa couplings 
$\lambda'_{1jk}$ is possible up to the kinematical limit of $\sim 300 \GeV$. 

Supersymmetric partners $\tilde{q}_L$ of left handed 
$u$-like squarks are produced preferentially in $e^+p$ collisions 
and most favourably via $\lambda'_{1j1}$.
In contrast, $e^-p$ collisions mainly produce partners $\tilde{q}_R$
of right-handed $d$-like quarks and most favourably via $\lambda'_{11k}$.
Squark decays via a $\lambda'$ coupling into $l + q$ final states 
dominate only at largest accessible masses, while elsewhere
squarks undergo mainly gauge decays into a quark and a gaugino-higgsino.
The $\tilde{q}_R$ decays involve a neutralino $\chi^0$ while 
$\tilde{q}_L$ decays dominantly proceed via a chargino $\chi^+$ in
a large portion of the MSSM parameter space.
The $\chi$'s, including the LSP, are generally unstable and their decay
chain involves the $\lambda'_{1jk}$ coupling.

In total, eight classes of event topologies are identified for $R$-parity
and gauge decays of squarks, with single or multi-leptons final states 
accompanied or not by missing transverse momenta.
A good experimental sensitivity is expected in each of these classes.
Thus, for an integrated luminosity of $500 \pbi$,
squarks can be searched for Yukawa couplings smaller than
the electromagnetic coupling up to masses of $\lesssim 270 \GeV$,
almost independently of the specific choice of MSSM parameter values.
Coupling values below the most stringent indirect limits can be
probed at $M_{\tilde{q}} = 150 \GeV$ for 
seven out of the nine possible $\lambda'_{1jk}$ couplings.

%===========================================================================

\vfill
\clearpage

%======================================================================
\begin{center} 
 {\large {\bf Appendix: Gaugino-higgsino mixing}}
\end{center} 
%======================================================================

Detailed expressions (and conventions) used for the mass mixing 
in the gaugino-higgsino sector are presented here for completeness. 

\vspace{0.3cm}

\noindent
{\bf Neutralino mass mixing:} \\
%    ----------------------
\noindent
Mass terms of the Lagrangian describing $SU(2)_L \times U(1)_Y$ 
neutral gauginos and higgsinos can be written as~: 
\begin{equation}
   \,\, {\cal{L}}_{m} = 
                 - 1/2 (\psi^0_i)^{T} Y^{ij} \psi^0_j + h.c. 
\end{equation}  
where the neutralino mass matrix in the basis
$  \psi^0_i =
   (-i\tilde{A},-i\tilde{W_{3}},\tilde{H}^{0}_{1},\tilde{H}^{0}_{2}) $
is given by~:
\begin{equation}
 Y=\left(
    \begin{array}{cccc}
     M_{1}  &  0     & -m_{Z}\sin\theta_{W}\cos\beta 
                                    &  m_{Z}\sin\theta_{W}\sin\beta \\
     0      &  M_{2} &  m_{Z}\sin\theta_{W}\cos\beta
                                    & -m_{Z}\cos\theta_{W}\sin\beta \\
    -m_{Z}\sin\theta_{W}\cos\beta      
            &  m_{Z}\cos\theta_{W}\cos\beta
                    &   0           & -\mu                          \\
     m_{Z}\sin\theta_{W}\sin\beta   
            & -m_{Z}\cos\theta_{W}\sin\beta
                    & -\mu          &  0                            \\
\end{array}
\right)
\end{equation}
%\end{center}
%
The number of free MSSM parameters is reduced by using a
GUT inspired relation between soft-breaking terms
$M_1$ and $M_2$, $ M_1 = \frac{5}{3} \tan^2 \theta_W M_2 $.

Neutralinos correspond to the mass eigenstates and are defined as
$ \chi^0_i = N^{ij} \psi^0_j$, with $N_{ij}$ being the unitary 
matrix which diagonalize $Y$. 
Finally, we make use of the matrix $N'$, which diagonalizes the
neutralino mass matrix expressed in the basis $(\tilde{\gamma},\tilde{Z})$
instead of $(\tilde{A}, \tilde{W}_3)$~:
$N'_{j1} = N_{j1} \cos \theta_W + N_{j2} \sin \theta_W$,
$N'_{j2} = -N_{j1} \sin \theta_W + N_{j2} \cos \theta_W$,
$N'_{j3} = N_{j3}$ and $N'_{j4} = N_{j4}$. 

\vspace{0.3cm}

\noindent
{\bf Chargino mass mixing:} \\
%    ----------------------
\noindent
The Lagrangian mass terms for winos and charged higgsinos are written
as~:
\begin{equation}
   {\cal{L}}_{m} = -\frac{1}{2} (\psi^{+}, \psi^{-})
       \left( \begin{array}{cc}
                  0    &X^{T} \\
                  X    &0     \\
              \end{array} \right)
       \left( \begin{array}{c}
                   \psi^{+}   \\
                   \psi^{-}   \\
        \end{array} \right)
    + \mbox{h.c}
\end{equation}
where~:
\begin{equation}
 X = \left(   \begin{array}{cc}
                   M_{2}                   & m_{W}\sqrt{2}\sin\beta  \\
                 m_{W}\sqrt{2}\cos\beta  & \mu
                \end{array}    \right)
\end{equation}
and $\psi^{+}_{j} = (-i \tilde{W}^{+}, \tilde{H}_{2}^{+})$,
$\psi^{-}_{j} = (-i \tilde{W}^{-}, \tilde{H}_{1}^{-})$.
%
% Les charginos (\'etats propres de masse)
% se d\'efinissent gr\^ace \`a une matrice $(4,4)$ qui s'\'ecrit par blocs
% \`a l'aide de deux matrices $(2,2)$ unitaires $U$ et $V$, dont on
% trouvera une
% d\'efinition claire dans~\cite{BARTLLEP2}~:
%
This mass matrix is diagonalized using two $(2,2)$ unitary matrices 
$U$ and $V$~\cite{BARTLLEP2}~: 
$ \chi^{+}_{i} = V^{ij} \psi^{+}_{j}$  and 
$ \chi^{-}_{i} = U^{ij} \psi^{-}_{j}$. 
Masses for these eigenstates are easily derived from the above 
$X$ matrix~:
\begin{equation}
\begin{array}{lll}
      M^2_{1,2} &=  &\frac{1}{2}(M_{2}^{2} +\mu^{2} +2m_{W}^{2} \\
  \mbox{} &\mbox{}  &\mp \left[ (M_{2}^{2}-\mu^{2})^{2}
+4m_{W}^{4}\cos^{2}2\beta \right. \\
  \mbox{} &\mbox{}  &\left.
+4m_{W}^{2}(M_{2}^{2}+\mu^{2}+2M_{2}\mu\sin2\beta) \right]^{1/2} \quad .\\
\end{array}
\end{equation}

\end{document}